\documentclass[journal,draftcls,onecolumn,12pt,twoside]{IEEEtran}
\usepackage{tabu}
\usepackage{multirow}
\usepackage{multicol}
\usepackage{array}
\usepackage{bm}
\usepackage{bbm}
\usepackage{subfigure}
\usepackage{stfloats}
\usepackage{setspace}
\usepackage{cite}
\usepackage{algorithm}
\usepackage{algorithmic}
\usepackage{subfigure}
\usepackage{textcomp}
\usepackage{float}
\usepackage{booktabs}
\usepackage{lineno}
\usepackage{caption}
\usepackage{rotating}
\usepackage{epstopdf}
\usepackage{amsmath,amssymb,amsfonts}
\usepackage{cases}
\usepackage[numbers,sort&compress]{natbib}
\usepackage{graphicx}
\usepackage{hyperref}
\usepackage{enumitem}
\usepackage{geometry}
\usepackage{color}
\usepackage[mathscr]{eucal}
\hypersetup{hypertex=true,
	colorlinks=true,
	linkcolor=black,
	anchorcolor=black,
	citecolor=black,
	urlcolor=blue}

\normalsize

\ifCLASSINFOpdf
\else
\fi

\begin{document}
	\title{IRS-Assisted RF-powered IoT Networks: System Modeling and Performance Analysis}
	\author{Hongguang~Sun,~\IEEEmembership{Member,~IEEE,}~Zelun~Zhao,~Hu~Cheng,
		~Jiangbin~Lyu,~\IEEEmembership{Member,~IEEE,} ~Xijun~Wang,~\IEEEmembership{Member,~IEEE,}
		~Yan~Zhang,~\IEEEmembership{Member,~IEEE,} and~Tony~Q.~S.~Quek,~\IEEEmembership{Fellow,~IEEE}
		
	\thanks{H. Sun, Z. Zhao, and H. Cheng are with the College of Information Engineering, Northwest A\&F University, Yangling 712100, China (e-mail: hgsun@nwafu.edu.cn; zhaozelun@nwafu.edu.cn; cheng\_gong\_mi@163.com). J. Lyu is with the School of Informatics, Xiamen University, China 361005 (e-mail: ljb@xmu.edu.cn). X. Wang is with the School of Electronics and Information Technology, Sun Yat-sen University, Guangzhou 510006, China (e-mail: wangxijun@mail.sysu.edu.cn). Y. Zhang is with the State Key Laboratory of Integrated Service Networks, Xidian University, Xi’an 710071, China (e-mail: yanzhang@xidian.edu.cn). T. Q. S. Quek is with the Information Systems Technology and Design Pillar, Singapore University of Technology and Design, Singapore 487372, and also with the National Cheng Kung University, Taiwan. (e-mail: tonyquek@sutd.edu.sg).}
	}
	\maketitle

\vspace{-4.5em}
\begin{abstract}
Emerged as a promising solution for future wireless communication systems, intelligent reflecting surface (IRS) is capable of reconfiguring the wireless propagation environment by adjusting the phase-shift of a large number of reflecting elements. To quantify the gain achieved by IRSs in the radio frequency (RF) powered Internet of Things (IoT) networks, in this work, we consider an IRS-assisted cellular-based RF-powered IoT network, where the cellular base stations (BSs) broadcast energy signal to IoT devices for energy harvesting (EH) in the charging stage, which is utilized to support the uplink (UL) transmissions in the subsequent UL stage. With tools from stochastic geometry, we first derive the distributions of the average signal power and interference power which are then used to obtain the energy coverage probability, UL coverage probability, overall coverage probability, spatial throughput \textcolor{blue}{and power efficiency,} respectively. With the proposed analytical framework, we finally evaluate the effect on network performance of key system parameters, such as IRS density, IRS reflecting element number, charging stage ratio, etc. Compared with the conventional RF-powered IoT network, IRS passive beamforming brings the same level of enhancement in both energy coverage and UL coverage, leading to the unchanged optimal charging stage ratio when maximizing spatial throughput. 
\end{abstract}

\vspace{-1em}

% \fi
\IEEEpeerreviewmaketitle

\vspace{-1em}
\section{Introduction}
\IEEEPARstart{W}{ith} the booming Internet of Things (IoT), the number of IoT devices is expected to reach tens of billions. To meet the vision of sustainable development of the IoTs, an efficient utilization of energy has been served as the principal issue. Recent advances in wireless energy harvesting (EH) technology, particularly radio frequency (RF) EH \cite{ref37}, have broken new ground to support IoT devices to collect energy from natural energy sources or ambient RF sources. It gave birth to the wireless powered communication networks (WPCNs) \cite{ref36} in which wireless nodes harvest energy from the base station (BS) RF signals and then transmit information by using the harvested energy. However, one of the major technical challenges in WPCN is the low efficiency of power transmission over long distances, resulting in the limited amount of energy harvested and poor network performance.  \\ \indent
In recent years, the application of intelligent reflect surface (IRS) technology has attracted extensive attention in wireless communications from both academia and industry. Served as a promising candidate technology for future wireless communications, IRS can smartly reconfigure the wireless channels by electronically configuring the absorption, reflection, refraction and phases via tuning the IRS reflecting elements or meta-atoms \cite{wu1}\cite{wu2}. Compared with the traditional active relaying or beamforming scheme, the passive reflection of IRSs consumes very limited energy by adopting the low-cost elements \cite{ref2}. Moreover, by properly adjusting the phase-shift of the IRS reflecting elements, the coverage probability, network throughput and energy efficiency can be greatly improved.\\ \indent
Using IRSs to boost the availability of WPCNs, IRS-assisted WPCN can be served as a potential network paradigm for future wireless communications. \textcolor{blue}{Understanding the gain achieved by IRSs in RF-powered IoT network is of great significance to speed up the application of IRSs and also help in the ingenious design of IRS-assisted WPCNs. The application of IRSs in WPCNs has been discussed in the very recent works \cite{phase4,ref22,ref23,ref24,ref38,ref39,ref40,R2}}. The authors in \cite{phase4,ref22,ref23,ref24,ref38,ref39,ref40,R2} considered the IRS-assisted WPCN scenarios under the NOMA scheme. Specifically, the authors in \cite{phase4} proved that utilizing different IRS phase-shifts for downlink (DL) EH and uplink (UL) packet transmission is not needed for the optimization of network throughput when NOMA is adopted. In \cite{ref22}, the sum-throughput was maximized by jointly optimizing the transmit power, transmit time and IRS beamforming. In \cite{ref23}, efficient schemes were proposed to maximize the UL sum rate by optimizing the resource allocation, reflection coefficients and beamformers. A self-sustainable IRS-empowered multi-user WPCN was considered in \cite{ref24}, where the IRS acts as a relay to enhance the network performance in both DL EH and UL information transmissions. In \cite{ref38}, the authors proposed to minimize the energy consumption of the hybrid access point (HAP) by jointly optimizing the transmit power, power-switching factor, time allocation, and IRS phase-shifting. The authors in \cite{ref39} studied an IRS-assisted multi-user multiple-input single-output (MISO) WPCN, in which the energy transmission time, the user transmit power, and the active and passive beamforming in DL energy collection and UL information transmissions were jointly optimized. In \cite{ref40}, the authors presented an IRS-assisted wireless powered caching network, in which the HAP maintains a local cache to store the popular contents for IoT devices. \textcolor{blue}{Given the imperfect CSI, the authors in \cite{R2} formulated a HAP transmit energy minimization problem by jointly optimizing HAP energy beamforming, receiving beamforming, etc. However, all the schemes designed in \cite{phase4,ref22,ref23,ref24,ref38,ref39,ref40,R2} were limited to the small-scale network scenario with only one BS and a single IRS, in which the distances between the BS,  the IRS and user equipments (UEs) are fixed and pre-determined. What's more, the location of IRS is assumed to be fixed, and the influence of interference is not considered. It's unknown whether these schemes are applicable to the practical multi-cell WPCNs.}  \\ \indent
\textcolor{blue}{In such a network, the spatial randomness of node locations, the time-varying channel fading, and the resulted complicated signal and interference distributions (caused by the reflection of IRSs) made it extremely difficult to evaluate the gain achieved by IRSs.} To address the aforementioned challenges, stochastic geometry has been explored over the past few years as a powerful tool in obtaining the average spatial system-level analysis of randomly deployed wireless networks, including cellular networks, and heterogeneous networks \cite{ref13,net1,net2,net3}. With this tool, the spatial randomness of node locations are modeled by some classical point processes, such as Poisson point process (PPP), and Poisson cluster process (PCP). Very recently, stochastic geometry has been introduced to analyze the performance of IRS-assisted cellular networks \cite{ref21}. Specifically, in \cite{ref21,st2,st5,st4,st6}, the authors characterized the DL coverage probability and spatial throughput of the IRS-assisted cellular network, which exhibited the gain achieved by IRSs on network throughput. In \cite{st2}, the authors demonstrated the superiority of equipping IRSs to the blockages. In \cite{st5}, the authors evaluated the performance of a millimeter wave (mmWave) network in which the average achievable rate was obtained by deriving the Laplace Transform of of the aggregated interference from all BSs and IRSs. In \cite{st4}, an analytical framework was proposed to analyze the performance of a DL IRS-assisted cellular network in which the coverage probability, ergodic capacity, and energy efficiency were derived. In \cite{st6}, the authors considered an IRS-aided multiple-input multiple-output (MIMO) network in which the outage probabilities, ergodic rates, spectral efficiency and energy efficiency were obtained. \textcolor{blue}{Very recently, IRSs have also been integrated with secure communication systems to defend against eavesdropping attacks and enhance the security performance \cite{R3,R4,R5}.}\\ \indent
\textcolor{blue}{To the best of our knowledge, there is no related works characterizing the performance of the large-scale IRS-assisted RF-powered IoT networks with tools from stochastic geometry. In such a network, the gain achieved by distributed IRSs in EH and information transfer is required to be quantified by considering the complicated network interference environment. }  \\ \indent
In this work, we consider an IRS-assisted cellular-based RF-powered IoT network, where IRSs help IoT devices harvest ambient RF energy from DL cellular transmissions in the charging stage, and assist UL transmissions in the subsequent UL stage. The main contributions of our work are summarized as follows:
\begin{itemize}
	\item 
	\textcolor{blue}{We propose an analytical framework for IRS-assisted cellular-based RF-powered IoT network, in which the locations of BSs, IoT devices, and IRSs are modeled by three independent PPPs. The IoT devices are battery-less and solely powered by the cellular networks.} Each time slot is assumed to be divided into two stages: i) charging stage, in which each IoT device harvests energy from the RF signals transmitted by BSs and reflected by the passive IRSs, and ii) UL stage, in which IoT devices with sufficient energy can transmit information to their associated BSs with the help of IRSs by taking into account the fractional power control strategy. \textcolor{blue}{For an IoT device, the passive beamforming scheme is adopted at its associated IRS in the whole time slot to accelerate the energy harvested by the IoT device and enhance the received signal power at its associated BS. }
	\item
	\textcolor{blue}{With the Gamma approximation method, we first characterize the signal power distributions in both charging and UL stages, and the interference distribution in the UL stages, based on which we further derive the tractable expressions of the energy coverage probability in the charging stage, the UL coverage probability in the UL stage, the overall coverage probability, the spatial throughput and the power efficency, respectively.} We then evaluate the influence of some key system parameters, such as BS density, IRS density and IRS reflecting element number on the derived performance metrics. 
	\item
	Compared with the conventional RF-powered IoT network, IRS passive beamforming significantly enhances the harvested signal power in the charging stage and the desired signal power in the UL stage, both of which are shown to scale with the number of IRS reflecting element $N$ in the same order of $O(N^2)$, while only slightly increasing the interference power. \textcolor{blue}{With the objective to maximize the network spatial throughput, the proposed framework allows to find the optimal charging stage ratio of the IRS-assisted RF-powered IoT network, which is shown to be the same as that without IRSs due to the same level of enhancement contributed by IRS passive beamforming on energy coverage and UL coverage.}
\end{itemize}   
\vspace{-0.5em}
\section{SYSTEM MODEL}
\vspace{-0.5em}
\subsection{Network Model}
\begin{figure}
	\centering
	\vspace{-2em}
	\subfigure[The process of IRS-assisted RF-powered IoT network during the whole time slot with $\tau$ and $T$ being the charging stage ratio, and the duration of one time slot, respectively. ]{
		\begin{minipage}{\linewidth} \label{IoTa}
			\centering
			\includegraphics[scale=0.45]{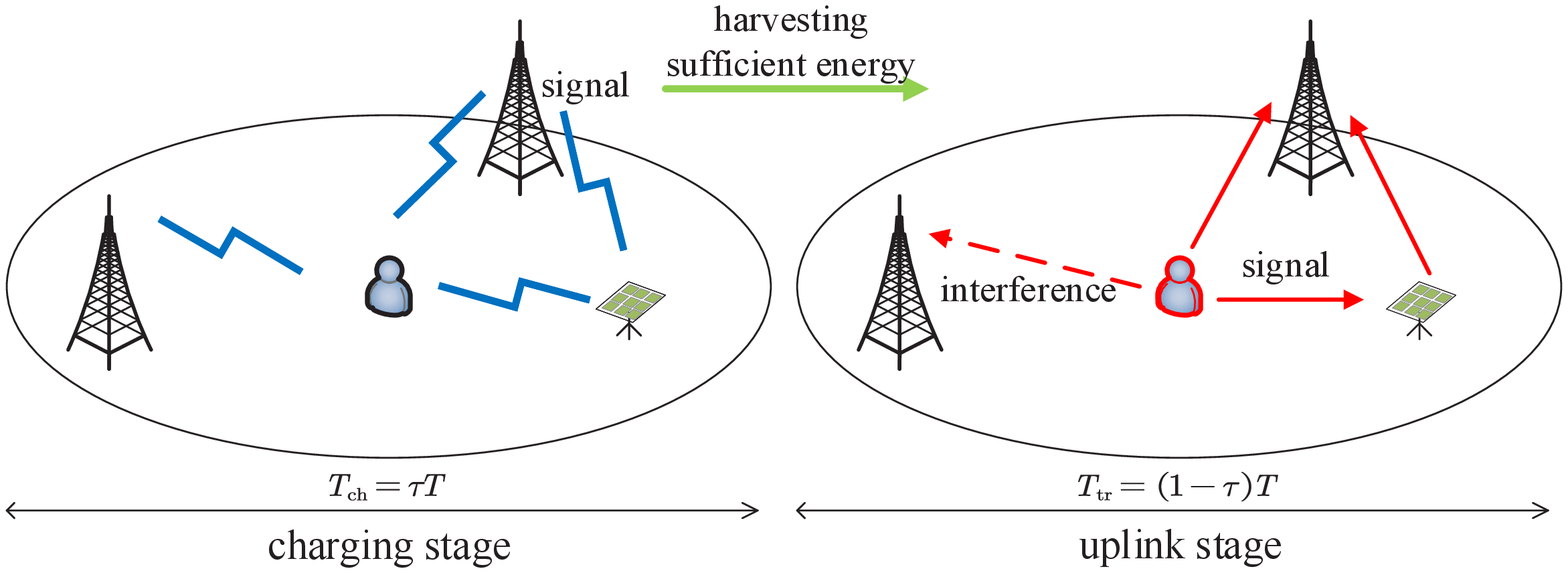}
		\end{minipage}
	}
	\centering
	\subfigure[IRS-assisted EH during the charging stage.]{
		\begin{minipage}[t]{0.45\linewidth}  \label{IoTb}
			\centering
			\includegraphics[scale=0.27]{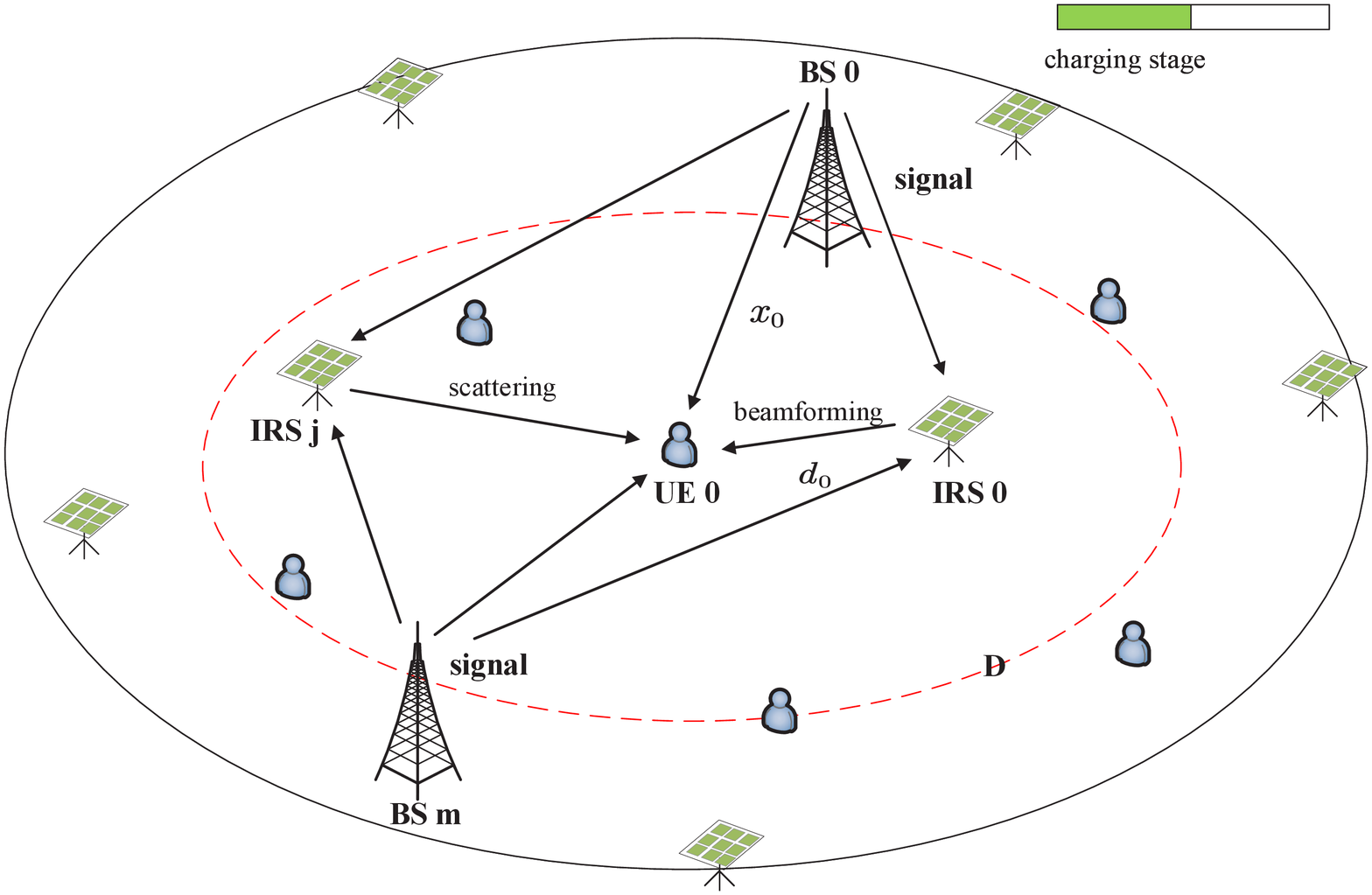}
		\end{minipage}
	}
	\centering
	\subfigure[IRS-assisted UL transmissions during the UL stage.]{
		\begin{minipage}[t]{0.46\linewidth}  \label{IoTc}
			\centering
			\includegraphics[scale=0.27]{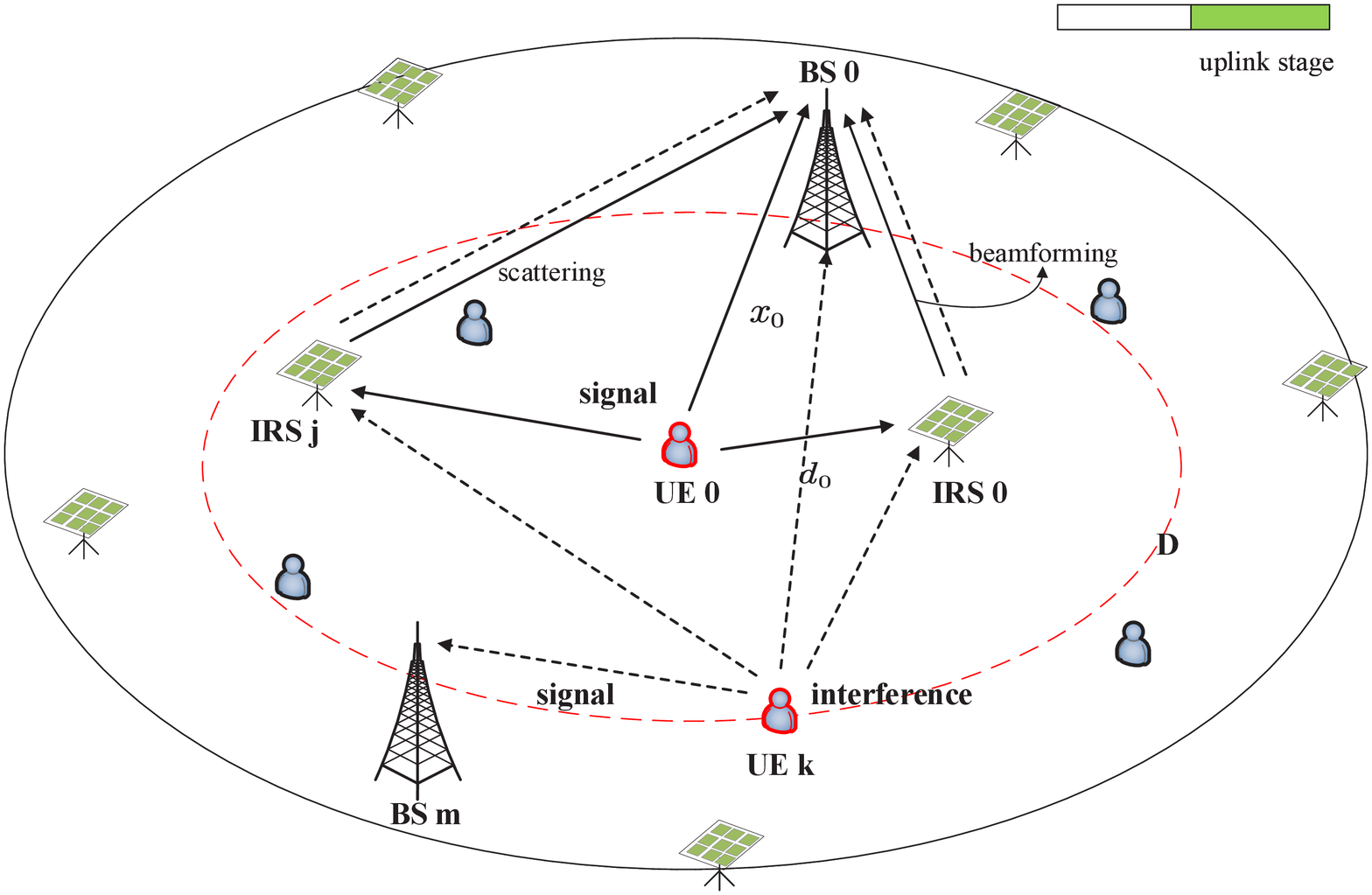}
		\end{minipage}
	}
	\centering
	\caption{An illustration of the IRS-assisted cellular-based RF-powered IoT network.}
	\label{fig:label}
\end{figure}

In this paper, we consider an IRS-assisted cellular-based IoT network where the IoT devices are battery-less and solely powered by the ambient RF energy only from cellular transmissions \cite{ref34}\cite{ref35}. We focus on the UL transmission, where the energy required by an IoT device to transmit in a given time slot should be harvested in the same time slot.  Specifically, all IoT devices are assumed to adopt the time-switching receiver architecture, with which an IoT device first harvests energy for a fraction of time and then transmits packets for the rest of time. As is shown in Fig. \ref{IoTa}, each time slot is divided into the charging stage and UL stage with durations $T_{\rm{ch}}=\tau T$ and $T_{\rm{tr}}=(1-\tau) T$, in which $T$ is the duration of each time slot, and $\tau$ is defined as the charging stage ratio. \textcolor{blue}{The antenna switching time between the two stages is neglected to facilitate the analysis.} The IoT device is assumed to have a supercapacitor with large charging and discharging rates to store and use the harvested energy during the same time slot. The remaining energy by the end of the time slot is assumed to be unavailable for the future transmissions taking into account the large leakage current of the supercapacitor \cite{joint}. \textcolor{blue}{We assume that both BSs and IoT devices are equipped with the single antenna.} The BSs are assumed to have the same height $H_{\rm{B}}$ with horizontal locations following a 2-dimensional (2D) PPP \( \Lambda_{\rm{B}}\) of density \( \lambda_{\rm{B}}\). Distributed IRSs are deployed to assist in both EH and UL transmissions, which are of the same height \(H_{\rm{I}}\) with the horizontal locations being modeled by a 2D PPP \( \Lambda_{\rm{I}}\) of density \( \lambda_{\rm{I}}\). IoT devices are scattered according to a PPP \(\Lambda_{\rm{u}}\) of density \(\lambda_{\rm{u}}\).  We adopt the orthogonal multiple access technology, such that only one IoT device within a cell can be active at any given time slot and sub-channel. \\ \indent
We assume open access and consider the nearest association policy, with which a given IoT device associates with the nearest BS, as well as the nearest IRS. According to Slivnyak’s theorem \cite{ref25}, it is sufficient to focus on a typical IoT device, referred to as UE 0 which is located at the origin and assumed to associate with BS 0 and IRS 0 (Fig. \ref{IoTb} and \ref{IoTc}). We define $x_m$, $d_j$ and  $r_{m,j}$ as the 2D distance from the BS $m$ to UE 0, the 2D distance from the IRS $j$ to UE 0, and the 2D distance from the BS $m$ to IRS $j$, respectively. Considering the limited reflective capability
of an IRS, we define a practical local region of radius $D$ with the typical UE 0 being the center, so that each IRS can only provide services for a limited number of UEs nearby. We denote IRSs within the local region by the set $\jmath\triangleq \left\{j\in\Lambda_{\rm{I}}|d_j \leq D\right\}$. Due to the densely deployed IRSs, we assume that there exists at least one IRS within set $\jmath$, and ignore the case $\jmath=\varnothing$. We consider the worst case where all IRSs are in working state no matter whether there are IoT devices to associate with, such that all IRSs can reflect the signal/interference from co-channel BSs to the typical UE 0. Considering the severe attenuation of wireless signals, the interference reflected by far away IRSs can be neglected. Thus, to simplify the analysis, we assume that only IRSs within set $\jmath$ contribute interference to the typical UE 0.   
\vspace{-0.5em} 
\subsection{IRS-Assisted EH and UL Transmission}
 As is shown in Fig.\ref{IoTb}, in the charging stage, all the BSs are served as RF chargers to broadcast energy signals. In Fig.\ref{IoTc}, during the UL stage, the IoT device with sufficient harvested energy transmits packets to its served BS. In both stages, all the IRSs within set $\jmath$ provide signal enhancement via random scattering or passive beamforming. Specially, to assist the EH and UL transmissions of UE 0, dedicated passive beamforming is adopted at its associated IRS 0, while other IRSs within $D$ only scatter the incident signal of BS 0.  \\ \indent
To support the UL transmission, the energy collected in the charging stage should reach a certain energy threshold $E_{\rm{min}}$, which is defined as the minimum energy required for the subsequent UL transmission. We adopt the fractional power control strategy for IoT devices where the transmit power of the typical IoT device is $\rho (x_0^2+H_{\text{B}}^2)^{\epsilon \alpha/2}$ with $x_0$ denoting the 2D distance between UE 0 and its serving BS, $\rho$ is the BS receiver sensitivity, and the symbol $\epsilon\in (0, 1]$ is defined as the power control parameter. According to the fractional power control strategy, $E_{\rm{min}}$ is given by 
\vspace{-0.5em}
\begin{equation} \label{Emin}
	\small
	E_{\rm{min}} =(1-\tau)T \rho (x_0^2+H_{\text{B}}^2)^{\frac{\epsilon \alpha}{2}}.
\end{equation}
Define $E_{\rm{h}}$ as the energy harvested by UE 0 during a charging stage, and UE 0 is allowed to transmit during the subsequent UL stage, as long as $E_{\text{h}} \geq E_{\rm{min}}$ is satisfied.
\vspace{-0.5em} 
\subsection{Channel Model}
For the sake of simplicity, we assume single antenna for both BSs and UEs, and define $N$ as the number of reflecting elements per IRS. We consider both the large-scale path loss and small-scale fading to characterize the channel model. For large-scale fading, the path loss exponent of BSs and IRSs is denoted by $\alpha$ with $\alpha>2$. We consider Rayleigh fading\footnote{ Rician fading or other channel fading models can also be suitable for the proposed analytical framework, in this paper we assume the worst-case propagation conditions for IRS, so as to obtain the lower bound of the achievable performance by IRS-assisted WPCN.} with unit power for the small-scale fading, which leads to an exponential channel power gain. Define $\sigma^2$ as the additive white Gaussian noise (AWGN). In the following, we discuss the channel model in charging stage and UL stage, respectively.
\subsubsection{Charging Stage}
In the charging stage, it is necessary to measure how much energy has been collected at the typical IoT device which depends upon the channel power gain. The baseline equivalent channels between BS $m$ and IRS $j$, between IRS $j$ and UE 0, and between BS $m$ and UE $0$ are represented by \(\boldsymbol{a}_{{\rm{i}},m}^{(j)}\triangleq\left[a_{{\rm{i}},m,1}^{(j)},\cdots,a_{{\rm{i}},m,N}^{(j)}\right]^\intercal\in \mathbb{C}^{N \times 1},\boldsymbol{a}_{\rm{r}}^{(j)}\triangleq\left[a_{\rm{r},1}^{(j)},\cdots,a_{{\rm{r}},N}^{(j)}\right]^\intercal \in \mathbb{C}^{N \times 1} \), and \( a_{{\rm{d}},m} \in \mathbb{C}\), respectively, where the symbols \(\left[\cdot\right]^\intercal\) and $\mathbb{C}$ denote the matrix transpose and the collection of complex numbers, respectively\footnote{\textcolor{blue}{The subscripts “i”, “r” and “d” represent the BS-to-IRS channel, the IRS-to-UE channel and direct BS-to-UE channel, respectively.}}. \textcolor{blue}{Define \( \phi^{(j)}\triangleq\left[\phi_1^{(j)},\cdots,\phi_N^{(j)}\right]\) and \(\boldsymbol{\Phi}^{(j)}\triangleq \rm{diag}\left\{\left[e^{\boldsymbol{\text{i}}\phi_1^{(j)}},\cdots,e^{\boldsymbol{\text{i}}\phi_N^{(j)}}\right]\right\}\) ($\boldsymbol{\text{i}}$ denotes the imaginary unit) as the phase-shift matrix of IRS $j$, where \(\phi_n^{(j)}\in\left[0,2\pi\right)\) is the phase-shift of the element $n$ of IRS $j$ on the incident signal. To achieve the maximal beamforming gain from IRS, we presume that the reflection coefficient of each reflecting element can achieve the unit amplitude \cite{phase1}.} Therefore, the cascaded BS-IRS-UE channel which is divided into BS-IRS transmission, IRS reflecting with beamforming, and IRS-UE transmission, is given by 
\vspace{-0.5em}
\begin{equation} \label{equ1}
	\small
	a_{{\rm{ir}},m}^{(j)}\triangleq\left[\boldsymbol{a}_{\rm{i},m}^{(j)}\right]^\intercal \boldsymbol{\Phi}^{(j)}\boldsymbol{a}_{\rm{r}}^{(j)}= \sum_{n=1}^{N}{a_{{\rm{i}},m,n}^{(j)}a_{{\rm{r}},n}^{(j)}e^{\boldsymbol{\text{i}}\phi_n^{(j)}}},m\in\Lambda_{\rm{B}}. 
\end{equation}  
The channel power gains between BS $m$-UE 0, between BS $m\in\Lambda_{\rm{B}}$ and the element $n$ of IRS $j$, and between the element $n$ of IRS $j$ and UE 0 are, respectively, is given by 
\vspace{-0.5em}
\begin{equation}\label{equ2}
	\small
	| a_{{\rm{d}},m}|^2\triangleq g_{{\rm{d}},m}\omega_{{\rm{d}},m},\quad
	| a_{{\rm{i}},m,n}^{(j)}|^2\triangleq g_{{\rm{i}},m}^{(j)}\omega_{{\rm{i}},m,n}^{(j)},\quad
	| a_{{\rm{r}},n}^{(j)}|^2\triangleq g_{\rm{r}}^{(j)}\omega_{{\rm{r}},n}^{(j)} 
\end{equation}
In (\ref{equ2}), $g_{{\rm{d}},m}=\beta\left(x_m^2 + H_{\rm{B}}^2\right)^{-\alpha/2}$, $g_{{\rm{i}},m}^{(j)}=\beta\left(r_{m,j}^2 + \left(H_{\rm{B}}-H_{\rm{I}}\right)^2\right)^{-\alpha/2}$ and $g_{\rm{r}}^{(j)}=\beta\left(d_j^2+H_{\rm{I}}^2\right)^{-\alpha/2}$ denote the corresponding average channel power gains,
  $\omega_{{\rm{d}},m}$, $\omega_{{\rm{i}},m,n}^{(j)}$ and $\omega_{{\rm{r}},n}^{(j)}$ are the small scale fading, and $\beta=\left(4\pi f_c/c\right)^{-2}$ is the average channel power gain at a reference distance of 1 m with $f_c$ being the carrier frequency, and $c=3.0\times10^8 \; (m/s)$ denoting the light speed. \textcolor{blue}{To reveal the effectiveness of IRSs in assisting the performance of commercial 5G networks operating on sub-6 GHz, we consider $f_{\text{c}} = 2$ GHz as the carrier frequency.}
\subsubsection{UL Stage}
For the UL stage, the baseband equivalent channels between UE $k$ and IRS $j$, between IRS $j$ and BS 0, and between UE $k$ and BS 0 are denoted by \(\boldsymbol{b}_{{\rm{i}},k}^{(j)}\triangleq\left[b_{\rm{i},k,1}^{(j)},\cdots,b_{{\rm{i}},k,n}^{(j)}\right]^\intercal\in \mathbb{C}^{N \times 1},\boldsymbol{b}_{\rm{r}}^{(j)}\triangleq\left[b_{\rm{r},1}^{(j)},\cdots,b_{{\rm{r}},N}^{(j)}\right]^\intercal \in \mathbb{C}^{N \times 1} \), and \( b_{{\rm{d}},k} \in \mathbb{C}\), respectively. Define \( \varphi^{(j)}\triangleq\left[\varphi_1^{(j)},\cdots,\varphi_N^{(j)}\right]\) and \(\boldsymbol{\Psi}^{(j)}\triangleq \rm{diag}\left\{\left[e^{\boldsymbol{\text{i}}\varphi_1^{(j)}},\cdots,e^{\boldsymbol{\text{i}}\varphi_N^{(j)}}\right]\right\}\) as the phase-shifting matrix of IRS $j$, of which \(\varphi_n^{(j)}\in\left[0,2\pi\right)\) represents the phase-shift of reflecting element $n$ on the incoming signal. Therefore, the cascaded UE-IRS-BS channel can be decomposed into three components: UE-IRS transmission, IRS reflecting, and IRS-BS transmission, expressed as
\vspace{-0.5em}
\begin{equation}\label{equ5}
	\small
	b_{{\rm{ir}},k}^{(j)}\triangleq\left[\boldsymbol{b}_{\rm{i},k}^{(j)}\right]^\intercal\boldsymbol{\Psi}^{(j)}\boldsymbol{b}_{\rm{r}}^{(j)}= \sum_{n=1}^{N}{b_{{\rm{i}},k,n}^{(j)}b_{{\rm{r}},n}^{(j)}e^{\boldsymbol{\text{i}}\varphi_n^{(j)}}} ,k\in\Lambda_{\rm{u}}.
\end{equation} \indent
The channel power gains between UE $k$ and BS 0, between UE $k$ and the element $n$ of IRS $j$, and between the element $n$ of IRS $j$ and BS 0 are, respectively, given by
\vspace{-1em}
\begin{equation}\label{equ6}
	| b_{{\rm{d}},k}|^2\triangleq f_{{\rm{d}},k}\xi_{{\rm{d}},k},\quad
	| b_{{\rm{i}},k,n}^{(j)}|^2 \triangleq f_{{\rm{i}},k}^{(j)}\xi_{{\rm{i}},k,n}^{(j)},\quad
	| b_{{\rm{r}},n}^{(j)}|^2\triangleq f_{\rm{r}}^{(j)}\xi_{{\rm{r}},n}^{(j)} ,
\end{equation}
where $f_{{\rm{d}},k}=\beta(y_k^2 + H_{\rm{B}}^2)^{-\alpha/2}$, $f_{{\rm{i}},k}^{(j)}=\beta(d_{k,j}^2 + H_{\rm{I}}^2)^{-\alpha/2}$ and $f_{\rm{r}}^{(j)}=\beta(r_j^2+(H_{\rm{B}}-H_{\rm{I}})^2)^{-\alpha/2}$ denote the corresponding average channel power gains, $\xi_{{\rm{d}},k}$, $\xi_{{\rm{i}},k,n}^{(j)}$ and $\xi_{{\rm{r}},n}^{(j)}$ are the small scale fading. The symbol $y_k$ denotes the horizontal distance from UE $k$ to the tagged BS,  and $D_k$ is the horizontal distance from the $k$-th device to its associated BS. According to the fractional power control strategy, the transmit power of UE $k$ is $\rho\left(D_k^2 + H_{\rm{B}}^2\right)^{\epsilon\alpha/2}$. To determine the transmit power of UE $k$, we obtain the distribution of $D_k$ conditioned on $y_k$ in \cite{ref13}, which is given by
\vspace{-0.5em}
\begin{equation}\label{equ95}
	\small
	f_{D_k}(r|y_k)=\frac{2\pi\lambda_{\rm{u}} r \exp(-\lambda_{\rm{u}} \pi r^2)}{1-\exp(-\pi\lambda_{\rm{u}} y_k^2)},0\leq r\leq y_k.
\end{equation}
\vspace{-2em}
\subsection{Channel Power Statistics}
To facilitate the performance analysis, we need to first derive the channel power statistics caused by the involvement of
IRSs in charging stage and UL stage, respectively.
\subsubsection{Charging Stage}
 In the charging stage, IoT devices harvest energy from the RF signals of BSs with the help of IRSs. With regards to the typical UE 0, considering the link BS $m$-IRS $j$-UE 0 in (\ref{equ1}), the reflected channel via element $n$ can be derived by
 \vspace{-1em}
\begin{equation}\label{equ9}
	\small
	\begin{aligned}
		a_{{\rm{ir}},m,n}^{(j)}\triangleq a_{{\rm{i}},m,n}^{(j)}a_{{\rm{r}},n}^{(j)}e^{\boldsymbol{\text{i}}\phi_n^{j}} = |a_{{\rm{i}},m,n}^{(j)}||a_{{\rm{r}},n}^{(j)}|e^{\boldsymbol{\text{i}}\left(\phi_n^{(j)}+\angle a_{{\rm{i}},m,n}^{(j)}+\angle a_{{\rm{r}},n}^{(j)}\right)},
	\end{aligned}
\end{equation} 
where the channel amplitude $|a_{{\rm{ir}},m,n}^{(j)}|\triangleq|a_{{\rm{i}},m,n}^{(j)}||a_{{\rm{r}},n}^{(j)}|$, and the cascade channel phase $\angle a_{{\rm{ir}},m,n}^{(j)}\triangleq\phi_n^{(j)}+\angle a_{{\rm{i}},m,n}^{(j)}+\angle a_{{\rm{r}},n}^{(j)}$. What's more, the amplitudes $|a_{{\rm{i}},m,n}^{(j)}|$ and $|a_{{\rm{r}},n}^{(j)}|$ follow the Rayleigh distribution with the scale parameters being $g^{(j)}_{{\rm{i}},m}$ and $g^{(j)}_{\rm{r}}$, respectively. As a result, each channel amplitude $|a_{{\rm{ir}},m,n}^{(j)}|$ is a double-Rayleigh RV with independent $|a_{{\rm{i}},m,n}^{(j)}|$ and $|a_{{\rm{r}},n}^{(j)}|$. We can derive the expectation and variance of $|a_{{\rm{ir}},m,n}^{(j)}|$ as
\vspace{-1em}
\begin{equation}\label{equ10}
	\small
	\mathbb{E}\left\{| a_{{\rm{ir}},m,n}^{(j)}|\right\}\triangleq\frac{\pi}{4}\sqrt{g_{{\rm{i}},m}^{(j)}g_{\rm{r}}^{(j)}} ,\quad
	{\rm{var}}\left\{| a_{{\rm{ir}},m,n}^{(j)}|\right\}\triangleq\left(1-\frac{\pi^2}{16}\right)g_{{\rm{i}},m}^{(j)} g_{\rm{r}}^{(j)}  .
\end{equation} \indent
According to the central limit theorem (CLT), the summation of $N$  independent and identically distributed (i.i.d.) RVs $X_1,X_2,\dots,X_N$, i.e., $Y=\sum_{n=1}^{N}X_n$ can be approximated as Gaussian distribution when $N$ is sufficiently large.
\\ \indent
With the IRS-customized channel estimation method \cite{ref29} \cite{ref30}, we can obtain the reflection cascade channel phase of each element on IRS 0. \textcolor{blue}{To be specific, IRS 0 can reconfigure each reflecting element's phase-shift by setting $\phi_n^{(0)}=-\angle(g_{{\rm{i}},0,n}^{(0)}g_{{\rm{r}},n}^{(0)}), (n=1,\dots,N)$, leading to the same phase of all $N$ reflected signals at UE 0,  where $g_{{\rm{i}},0,n}^{(0)}$ and $g_{{\rm{r}},n}^{(0)}$ is defined as the average channel power gains of between BS 0 and the element $n$ of IRS $j$ and between the element $n$ of IRS $j$ and UE 0.} As a result, the signal at the typical UE 0 is the sum of $N$ reflected signals. Thus, the BS 0-IRS 0-UE 0 channel can be derived by
\vspace{-0.5em}
\begin{equation}\label{equ12}
	\small
	| a_{{\rm{ir}},0}^{(0)}|=| \boldsymbol{a}_{{\rm{i}},0}^{(0)}|^\intercal| \boldsymbol{a}_{\rm{r}}^{(0)}|=\sum_{n=1}^{N}{| a_{{\rm{i}},0,n}^{(0)}|| a_{{\rm{r}},n}^{(0)}|},
\end{equation}
which is the summation of $N$ i.i.d. double-Rayleigh RVs, the channel amplitude is approximated
to follow the Gaussian distribution
\vspace{-0.5em}
\begin{equation}\label{equ13}
	\small
	|a_{{\rm{ir}},0}^{(0)}| \overset{\text{approx.}}{\sim}\mathcal{N}\left(N\frac{\pi}{4}\sqrt{g_{{\rm{i}},0}^{(0)}g^{(0)}_{{\rm{r}}}},N(1-\frac{\pi^2}{16})g_{{\rm{i}},0}^{(0)}g_{{\rm{r}}}^{(0)}\right).
\end{equation} \indent
Therefore, the average signal power of the cascaded BS 0-IRS 0-UE 0 channel can be computed by the second moment of $|a_{{\rm{ir}},0}^{(0)}|$ which is obtained by
\vspace{-0.5em}
\begin{equation}\label{equ14}
	\small
	g_{{\rm{ir}},0}^{(0)}\triangleq \mathbb{E}\left\{ | a_{{\rm{ir}},0}^{(0)}|^2 \right\} = \left[\frac{\pi^2}{16}N^2 + \left(1-\frac{\pi^2}{16}\right)N\right]g_{{\rm{i}},0}^{(0)}g_{\rm{r}}^{(0)},
\end{equation}
where $ G_{\rm{sc}}\triangleq[\frac{\pi^2}{16}N^2+(1-\frac{\pi^2}{16})N] $ denotes the beamforming coefficient of $g_{{\rm{i}},0}^{(0)}g_{\rm{r}}^{(0)}$, growing with $N$ in the order of $O(N^2)$. \\ \indent
\textcolor{blue}{With regards to any other IRS $j\in\jmath$ without providing beamforming for UE 0, it scatters the incident signal from BS $m$ randomly, leading to a uniformly random phase $\angle a_{{\rm{ir}},m,n}^{(j)}$ based on $\angle a_{{\rm{i}},m,n}^{(j)}$ and  $a_{{\rm{r}},n}^{(j)}$.} Therefore, for each reflecting element $N$, the cascaded channel $a_{{\rm{ir}},m,n}^{(j)}$ has zero mean and independent in-phase and quadrature-phase components each with variance $\frac{1}{2}a_{{\rm{i}},m,n}^{(j)}a_{{\rm{r}},n}^{(j)}$. According to the CLT, for a practically large N, we can approximate both the in-phase and quadrature-phase of $a_{{\rm{ir}},m}^{(j)}=\sum_{n=1}^{N}a_{{\rm{ir}},m,n}^{(j)}$ by Gaussian distribution $\mathcal{N}(0,\frac{1}{2}N a_{{\rm{i}},m}^{(j)}a_{\rm{r}}^{(j)})$. As a result, we use the following CSCG distribution to approximate the BS $m$-IRS $j$-UE 0 channel, given by 
\vspace{-0.5em}
\begin{equation}\label{equ15}
	\small
	a_{{\rm{ir}},m}^{(j)} \overset{\text{approx.}}{\sim} \mathcal{CN}\left(0,Ng_{{\rm{i}},m}^{(j)}g_{\rm{r}}^{(j)}\right).
\end{equation}\indent
Thus, the average channel power of BS $m$-IRS $j$-UE 0 link can be derived by
\vspace{-0.5em}
\begin{equation}\label{equ16}
	\small
	g_{{\rm{ir}},m}^{(j)}\triangleq\mathbb{E}\left\{| a_{{\rm{ir}},m}^{(j)}|^2\right\}=Ng_{{\rm{i}},m}^{(j)}g_{\rm{r}}^{(j)}.
\end{equation}
\vspace{-2.5em}
\subsubsection{UL Stage}
In the UL stage, IoT devices that harvest sufficient energy transmit packets to their serving BSs with the help of IRSs. With regards to the tagged BS 0, considering the link UE $k$-IRS $j$-BS 0 in (\ref{equ5}), the reflected channel via element $n$ can be derived by 
\vspace{-0.5em}
\begin{equation}\label{equ17}
	\begin{aligned}
		\small
		b_{{\rm{ir}},k,n}^{(j)}\triangleq b_{{\rm{i}},k,n}^{(j)}b_{{\rm{r}},n}^{(j)}e^{\boldsymbol{\text{i}}\varphi_n^{j}} = |b_{{\rm{i}},k,n}^{(j)}||b_{{\rm{r}},n}^{(j)}|e^{\boldsymbol{\text{i}}\left(\varphi_n^{(j)}+\angle b_{{\rm{i}},k,n}^{(j)}+\angle b_{{\rm{r}},n}^{(j)}\right)},
	\end{aligned}
\end{equation} 
where the channel amplitude $|b_{{\rm{ir}},k,n}^{(j)}|\triangleq|b_{{\rm{i}},k,n}^{(j)}||a_{{\rm{r}},n}^{(j)}|$, and the cascade channel phase $\angle b_{{\rm{ir}},k,n}^{(j)}\triangleq\varphi_n^{(j)}+\angle b_{{\rm{i}},k,n}^{(j)}+\angle b_{{\rm{r}},n}^{(j)}$. What's more, the amplitudes $|b_{{\rm{i}},k,n}^{(j)}|$ and $|b_{{\rm{r}},n}^{(j)}|$ follow the Rayleigh distribution with the scale parameters being $f_{{\rm{i}},k}^{(j)}$ and $f_{r}^{(j)}$, respectively. Similarly, the amplitude of the UE 0-IRS 0-BS 0 channel is derived as
\vspace{-0.5em}
\begin{equation}\label{equ18}
	\small
	| b_{{\rm{ir}},0}^{(0)}|=| \boldsymbol{b}_{{\rm{i}},0}^{(0)}|^\intercal| \boldsymbol{b}_{\rm{r}}^{(0)}|=\sum_{n=1}^{N}{| b_{{\rm{i}},0,n}^{(0)}|| b_{{\rm{r}},n}^{(0)}|},
\end{equation}
which for large $N$ is approximated as Gaussian distribution.
\vspace{-0.5em}
\begin{equation}\label{equ19}
	\small
	| b_{{\rm{ir}},0}^{(0)}|\overset{\text{approx.}}{\sim}\mathcal{N}\left(N \frac{\pi}{4}\sqrt{f_{{\rm{i}},0}^{(0)}f_{\rm{r}}^{(0)}},N\left(1-\frac{\pi^2}{16}\right)f_{{\rm{i}},0}^{(0)}f_{\rm{r}}^{(0)}\right).
\end{equation}\indent
For a sufficiently large $N$, we can use the CSCG distribution to approximate the UE $k$-IRS $j$-BS 0 channel, which is obtained by
\vspace{-1em}
\begin{equation}\label{equ20}
	\small
	b_{{\rm{ir}},k}^{(j)} = \sum_{n=1}^{N}{b_{{\rm{ir}},k,n}^{(j)}} \overset{\text{dist.}}{\sim} \mathcal{CN}\left(0,Nf_{{\rm{i}},k}^{(j)}f_{\rm{r}}^{(j)}\right).
\end{equation}\indent
Therefore, the UE 0-IRS 0-BS 0 and UE $k$-IRS $j$-BS 0 average channel power are, respectively, given by
\vspace{-1em}
\begin{equation}\label{equ21}
	\small
	f_{{\rm{ir}},0}^{(0)}\triangleq \mathbb{E}\left\{ | b_{{\rm{ir}},0}^{(0)}|^2 \right\}=G_{\rm{sc}}f_{{\rm{i}},0}^{(0)}f_{\rm{r}}^{(0)},\quad
	f_{{\rm{ir}},k}^{(j)}\triangleq\mathbb{E}\left\{| b_{{\rm{ir}},k}^{(j)}|^2\right\}=Nf_{{\rm{i}},k}^{(j)}f_{\rm{r}}^{(j)}.
\end{equation}
where $ G_{\rm{sc}}\triangleq[\frac{\pi^2}{16}N^2+(1-\frac{\pi^2}{16})N] $ denotes the beamforming coefficient of $f_{{\rm{i}},0}^{(0)}f_{\rm{r}}^{(0)}$, growing with $N$ in the order of $O(N^2)$.
\vspace{-1.6em}
\subsection{Performance Metrics}
\vspace{-0.1em}
In this subsection, we detail the performance metrics considered in this work. To be specific, we consider the following metrics: energy coverage probability in the charging stage, UL coverage probability during the UL stage, overall coverage probability and network spatial throughput. 
\subsubsection{Energy Coverage Probability}
Define $P_{\rm{t}}$ as the BS transmit power. Under the condition that UE 0 is associated with the nearest BS 0, the amount of energy harvested from BS 0 is composed of the following three parts: from the direct transmission of BS 0, from the reflected transmission through IRS 0  performing beamforming, and from the reflected transmission through other IRSs $\in \jmath$ performing random scattering. Specifically, the harvested signal power from BS 0 is obtained by
\vspace{-1em}
\begin{equation}\label{equ23}
	\small
	S_{\rm{dr}}\triangleq P_{\rm{t}}\cdot | a_{{\rm{d}},0} + \sum_{j\in \jmath}{a_{{\rm{ir}},0}^{(j)}}|^2.
\end{equation}
Similarly, the amount of harvested signal power from BS $m\in\Lambda_{\rm{B}}\backslash \left\{ 0\right\}$ consists of the following two parts: from the direct transmission of BS $m$, and from the reflected transmission through all IRSs $\in \jmath$ performing random scattering. To be specific, the harvested signal power from BS $m$ can be derived as
\vspace{-1em}
\begin{equation}\label{equ24}
	\small
	S_{\rm{id}}\triangleq P_{\rm{t}} \cdot\sum_{m\in\Lambda_{\rm{B}}\backslash\left\{ 0\right\}}{| a_{{\rm{d}},m} + \sum_{j\in\jmath}{a_{{\rm{ir}},m}^{(j)}}|^2}.
\end{equation} \indent
In summary, the overall amount of harvested energy at UE 0 is given by 
\vspace{-0.5em}
\begin{equation}\label{equ25}
	\small
	E_{h} = \tau T\eta \left(S_{\rm{dr}} + S_{\rm{id}}\right)\quad \rm{Joules},
\end{equation}
where $\eta<1$ indicates the RF energy conversion efficiency. \textcolor{blue}{To simplify the analysis, we adopt the linear energy-harvesting model as in \cite{ref22} and assume that the input power to the circuit of receiver belongs to the low power regime without saturation of EH. Note that our work can be extended to the non-linear EH models by considering the saturation regime.} To show the effectiveness of EH, we define the energy coverage probability during the charging stage, which is given by 
\vspace{-0.5em}
\begin{equation}\label{equ27}
	\small
	\bold{P}_{\rm{en}} = \mathbb{E}\left[\mathbb{I}\left( E_{\rm{h}} \geqslant E_{\rm{min}}\right)\right],
\end{equation}
where $E_{\rm{min}} = (1-\tau)T\rho (x_0^2+H_{\text{B}}^2)^{\epsilon \alpha/2}$ in (\ref{Emin}) with $\mathbb{I}(\cdot)$ being the indicator function. 
\subsubsection{UL Coverage Probability}
According to the UL fractional power control strategy, the transmit power of UE 0 and UE $ k\in\Lambda_{\rm{u}}'\backslash\left\{0\right\}$ is, respectively,  given by $P_0=\rho (x_0^2+H_{\text{B}}^2)^{\epsilon\alpha/2}$ and $P_k= \rho (D_k^2+H_{\text{B}}^2)^{\epsilon\alpha/2}$. The desired signal received at the tagged BS 0 is composed of direct signal from UE 0 and the reflected signal of UE 0 via all IRSs $\in \jmath$ (i.e., performing beamforming and random scattering). Therefore, we derive received signal power by 
\begin{equation}\label{equ28}
	\small
	S_{\rm{UL}}\triangleq \rho (x_0^2+H_{\rm{B}}^2)^{\frac{\epsilon\alpha}{2}}\cdot | b_{{\rm{d}},0} + \sum_{j\in \jmath}{b_{{\rm{ir}},0}^{(j)}}|^2.
\end{equation}\indent
After the charging stage, the active IoT devices that are harvesting sufficient energy and transmitting on the same time-frequency resource block with UE 0 can be approximated to follow a ${\rm{PPP}} \sim \Lambda_{\rm{u}}^{'}$ of density $\lambda_{\rm{u}}^{'} = \bold{P}_{\rm{en}}\lambda_{\rm{B}}$ with $\bold{P}_{\rm{en}} = \mathbb{E}\left[\mathbb{I}\left( E_{\rm{h}} \geqslant E_{\rm{min}}\right)\right]$ given by (\ref{equ27}). Therefore, the overall interference received by BS 0 is given by
\begin{equation}\label{equ29}
	\small
	I\triangleq\sum_{k\in\Lambda_{\rm{u}}'\backslash\left\{0\right\}}{I_k}=\sum_{k\in\Lambda_{\rm{u}}'\backslash\left\{0\right\}}{\rho (D_k^2+H_{\rm{B}}^2)^{\frac{\epsilon\alpha}{2}}|b_{{\rm{d}},k} + \sum_{j\in\jmath}{b_{{\rm{ir}},k}^{(j)}}|^2}.
\end{equation}\indent
The received SINR at BS 0 is given by
\begin{equation}\label{equ30}
	\begin{aligned}
	\small
	{\gamma}\triangleq\frac{S_{\rm{UL}}}{I+\sigma^2}=\frac{\rho (x_0^2+H_{\rm{B}}^2)^{\frac{\epsilon\alpha}{2}}\cdot | b_{{\rm{d}},0} + \sum_{j\in \jmath}{b_{{\rm{ir}},0}^{(j)}}|^2}{\sum_{k\in\Lambda_{\rm{u}}'\backslash\left\{0\right\}}{\rho (D_k^2+H_{\rm{B}}^2)^{\frac{\epsilon\alpha}{2}}|b_{{\rm{d}},k} + \sum_{j\in\jmath}{b_{{\rm{ir}},k}^{(j)}}|^2} + \sigma^2}.
	\end{aligned}
\end{equation} \indent
The UL coverage probability is actually a conditional coverage probability under the condition that the typical UE 0 has harvested sufficient energy is given by
\begin{equation}\label{equ31}
	\small
	\bold{P}_{\rm{UL}}\triangleq\mathbb{P}\left\{{\gamma}\geq{\overline{\gamma}}\lvert E_{\rm{h}} \geq E_{\rm{min}}\right\},
\end{equation}
where $\overline{\gamma}$ is the SINR threshold. 
\subsubsection{Overall Coverage Probability }
By definition, the overall coverage probability is defined as the joint probability of the aforementioned two events, which can be expressed by
\begin{equation}\label{equ32}
	\small
	\bold{P}_{\rm{cov}}\triangleq\mathbb{E}\left[ \mathbb{I}\left(E_{\rm{h}}\geq E_{\rm{min}}\right)\mathbb{I}\left( {\gamma}\geq{\overline{\gamma}}\right)\right].
\end{equation} 
\subsubsection{Spatial Throughput}
Based on the overall coverage probability, the average spatial throughput can be obtained as 
\begin{equation}\label{equ82}
	\small
	\begin{aligned}
		&\nu =(1-\tau)\lambda_{\rm{u}}'\mathbb{E}\left[\log\left(1+\overline{\gamma}\right)\mathbb{I}\left(E_{\rm{h}}\geq E_{\rm{min}}\right)\times\mathbb{I}\left(\gamma\geq \overline{\gamma}\right)\right]=(1-\tau)\bold{P}_{\rm{en}}\lambda_{\rm{u}}\log\left(1+\overline{\gamma}\right)\bold{P}_{\rm{cov}},
	\end{aligned}
\end{equation}
where $\bold{P}_{\rm{en}}$ and $\bold{P}_{\rm{cov}}$ are given by (\ref{equ27}) and (\ref{equ32}), the expressions of which will be derived in the following section.
\subsubsection{Power Efficiency}
\textcolor{blue}{ We consider the following two performance metrics to illustrate the power efficiency: energy harvesting efficiency (EHE) \cite{r30} in the charging stage and energy efficiency (EE) \cite{r31} in the UL stage, where the former is defined as the amount of power scavenged by all the IoT devices per unit power consumed by the BS, while the latter refers to the number of bits transmitted by an IoT device per Joule. 
The EHE can be expressed as
	\begin{equation}\label{ehef}
		\small
		\mathscr{EHE}=\frac{\mathbb{E}\{E_{\text{h}}\}\lambda_{\rm{u}}}{\tau T\lambda_{\text{B}}(P_{\text{cb}}+\frac{1}{\eta_{\text{b}}}P_{\text{t}})},
	\end{equation}
	where $\mathbb{E}\{E_{h}\}$ is the average amount of energy harvested by the typical UE 0 during the charging stage, $\eta_{\text{b}}\in\left(0,1\right]$ represents the efficiency of the power amplifier efficiency of a BS, $P_{\text{cb}}$ and $P_{\text{t}}$ denote the static power and transmit power of a BS, respectively. The EE of the typical UE 0 can be expressed as 
	\begin{equation}\label{eef}
		\small
		\mathscr{EE}=\frac{(1-\tau)T \log\left(1+\overline{\gamma}\right)\bold{P}_{\rm{cov}}}{(P_{\text{cu}}+{\frac{1}{\eta_{\text{u}}}\mathbb{E}\{P_\text{0}\}})(1-\tau)T},
	\end{equation}
	where the numerator denotes the number of bits transmitted within the UL stage, and the denominator represents the average energy consumption of the typical UE 0 with $\mathbb{E}\{P_0\}$ being the transmit power,  $\eta_{\text{u}}\in\left(0,1\right]$ being the efficiency of the power amplifier efficiency of an IoT device, and $P_{\text{cu}}$ being the static power of typical UE 0, respectively. }
\vspace{-0.5em}
\section{PERFORMANCE ANALYSIS}
In this section, we first study the signal power distribution in both stages, and then characterize the interference distribution in the UL stage. The aforementioned performance metrics are finally obtained.
\vspace{-1.5em}
\subsection{Signal Power Distribution}
The signal power distribution is related to not only the charging stage, but also the UL stage. To be specific, in the charging stage, all BSs broadcast energy signals which are harvested by IoT devices. While in the UL stage, for BS 0, only that originated from UE 0 is the desired signal.  \\ \indent
In the charging stage, the energy harvested from BSs is contributed from both the direct signal and the reflected signals via all IRSs within set $\jmath$. In the following, we first derive the signal power distribution from BS 0 conditioned on $x_0$ and $d_0$. The overall conditional signal can be expressed as $a_{{\rm{d}},0}+\sum_{j\in\jmath}a_{{\rm{ir}},0}^{(j)}$, the summation of a RV of Rayleigh distribution and N RVs of Gaussian distribution. Because of the difficulty in deriving the exact signal power distribution from BS 0, we use the Gamma distribution \cite{ref33} as an alternative. Specifically, we use Gamma distribution to approximate the distribution of $S_{\rm{dr}}$, given by
\begin{equation}\label{equ34}
	\small
	S_{\rm{dr}}|_{d_0,x_0} \overset{\text{approx.}}{\sim} \Gamma[k_{\rm{dr}},\theta_{\rm{dr}}],
\end{equation}
where $k_{\rm{dr}}$ and $\theta_{\rm{dr}}$ represent the shape parameter and scale parameter, respectively. With the moment matching technique \cite{ref31}, we have
\begin{equation}\label{equ35}
	\small
	k_{\rm{dr}}\triangleq\frac{(\mathbb{E}\{S_{\rm{dr}}\}|_{d_0,x_0})^2}{{\rm{var}}\{S_{\rm{dr}}\}|_{d_0,x_0}},\quad \theta_{\rm{dr}}\triangleq\frac{{\rm{var}}\{S_{\rm{dr}}\}|_{d_0,x_0}}{\mathbb{E}\{S_{\rm{dr}}\}|_{d_0,x_0}}.
\end{equation} 
To obtain $k_{\rm{dr}}$ and $\theta_{\rm{dr}}$, we derive the first and second moments of $S_{\rm{dr}}$ conditioned on $d_0$ and $x_0$.
\vspace{-0.5em}
\begin{equation}\label{equ36}
	\small
	\begin{aligned}
		&\mathbb{E}\left\{ S_{\rm{dr}}\right\}\vert_{d_0,x_0} \approx  P_{\text{t}}\cdot( \mathbb{E}\left\{ | a_1|^2\right\}|_{d_0,x_0} + \mathbb{E}\left\{ | a_2|^2\right\}|_{d_0,x_0}),
	\end{aligned}
\end{equation}
\begin{equation}\label{equ37}
	\small
	\begin{aligned}
		&\mathbb{E}\left\{ S_{\rm{dr}}^2\right\}|_{d_0,x_0} =P_{\text{t}}\cdot( \mathbb{E}\{ |a_1|^4\}|_{d_0,x_0} + \mathbb{E}\{ |a_2|^4\}|_{d_0,x_0} + 4\mathbb{E}\{ |a_1|^2\}|_{d_0,x_0}\mathbb{E}\{ |a_2|^2\}|_{d_0,x_0}).
	\end{aligned}
\end{equation}\indent
The first two moments of $S_{\rm{dr}}|_{d_0,x_0}$ can be derived by the first two moments of $a_1$, $a_2$, $| a_1|^2$, $| a_2|^2$. With the similar method proposed in Appendix B of \cite{ref21} and  the approximation of $r_{0,j}\approx x_0$,  $g_{{\rm{i}},0}^{(j)}\approx g_{{\rm{d}},0}$ for $j\in\jmath$, the first two moments of  $a_1$, $a_2$, $| a_1|^2$, $| a_2|^2$ can be derived by
\begin{equation}\label{a1}
	\small
	\mathbb{E}\{a_1\}|_{d_0,x_0}=0,\quad\mathbb{E}\{a_1^2\}|_{d_0,x_0}=0,\quad
	\mathbb{E}\{a_2\}|_{d_0,x_0}=0,\quad\mathbb{E}\{a_2^2\}|_{d_0,x_0}=0,
\end{equation}
\begin{equation}\label{a1^2}
	\small
	\mathbb{E}\{ |a_1|^2\}|_{d_0,x_0}\approx g_{{\rm{d}},0}( 1 + G_{\rm{sc}}g_{\text{r}}^{(0)} + N\frac{\pi}{4}\sqrt{\pi g_{\text{r}}^{(0)}}),
\end{equation}
\begin{equation}\label{a1^4}
	\small
	\begin{aligned}
		\mathbb{E}\{|a_1|^4\}|_{d_0,x_0}&\approx	\left[g_{{\rm{d}},0}\right]^2\left[2 + \frac{3}{4}\pi^{\frac{3}{2}}N\sqrt{g_{\text{r}}^{(0)}} +  6G_{\rm{sc}} g_{\text{r}}^{(0)}+2\sqrt{\pi}\left(\frac{\pi^3N^3}{64}+\frac{3\pi N^2(1-\frac{\pi^2}{16})}{4}\right)\right.\\&\left.\times\left[g_{\text{r}}^{(0)}\right]^{\frac{3}{2}}+\left(\frac{\pi^4 N^4}{256} + \frac{3\pi^2 N^3(1-\frac{\pi^2}{16})}{8} + 3N^2(1-\frac{\pi^2}{16})^2\right)\left[g_{\text{r}}^{(0)}\right]^2\right],
	\end{aligned}
\end{equation}
\begin{equation}\label{a2^2}
	\small
	\mathbb{E}\{|a_2|^2\}|_{d_0,x_0}\approx Ng_{{\rm{d}},0}E_{\rm{S1}}(d_0), \quad \mathbb{E}\{|a_2|^4\}|_{d_0,x_0}\approx 2N^2[g_{{\rm{d}},0}]^2E_{\rm{S3}}(d_0),
\end{equation}
where $g_{\rm{d,0}}$ and $g_{\rm{r}}^{(0)}$ denote the average channel power gain of the BS 0-UE 0 and IRS 0-UE 0 link, i.e.,
\begin{equation}\label{gdx}
	\small
	g_{\rm{d,0}}\triangleq\beta\left( x_0^2 + H_{\rm{B}}^2\right)^{-\alpha/2},\quad
	g_{\rm{r}}^{(0)}\triangleq\beta\left( d_0^2 + H_{\rm{I}}^2\right)^{-\alpha/2},
\end{equation} 
and $E_{\rm{S1}}$ denotes the expectation of $\sum_{j\in\jmath\backslash\{0\}}g_{r}^{(j)}$ which is given by
\begin{equation}\label{S1}
	\small
	\begin{aligned}
			E_{\rm{S1}}(d_0)=\frac{2\pi\lambda_{\rm{I}}\beta}{\alpha-2}\left[ \left( d_0^2 + H_{\rm{I}}^2\right)^{1-\frac{\alpha}{2}} - \left( D^2 + H_{\rm{I}}^2\right)^{1-\frac{\alpha}{2}}\right],
	\end{aligned}
\end{equation}
and
\vspace{-0.5em}
\begin{equation}\label{S3}
	\small
	\begin{aligned}
		E_{\rm{S3}}\triangleq\mathbb{E}\left\{\left(\sum_{d_0<d_j \leq D}{g_{\rm{r}}(d_j)}\right)^2\right\}=(E_{\rm{S1}}(d_0))^2 + E_{\rm{S2}}(d_0),
	\end{aligned}
\end{equation}
with $E_{\rm{S2}}\left(d_0\right)$ given by
\vspace{-1.5em}
\begin{equation}\label{S2}
	\small
	\begin{aligned}
		E_{\rm{S2}}\left(d_0\right) = \frac{\pi\lambda_{\rm{I}}\beta^2}{\alpha-1}\left[(d_0^2 + H_{\rm{I}}^2)^{1-\alpha} - (D^2 + H_{\rm{I}}^2)^{1-\alpha}\right].
	\end{aligned}
\end{equation} \indent
Therefore, we can derive the first and second moments of $S_{\rm{dr}}$ conditioned on $d_0$ and $x_0$. The variance is derived by \( {\rm{var}}\left\{ S_{\rm{dr}}\right\}\vert_{d_0,x_0}\triangleq E\left\{S_{\rm{dr}}^2\right\}\vert_{d_0,x_0}-\left(E\left\{S_{\rm{dr}}\right\}\vert_{d_0,x_0}\right)^2 \). By substituting the above expressions into (\ref{equ35}), we derive the shape parameter $k_{\rm{dr}}$ and $\theta_{\rm{dr}}$, respectively.  \\ \indent
 Based on (\ref{equ36}), (\ref{a1^2}) and (\ref{a2^2}), the conditional mean signal power is the product of $g_{{\rm{d}},0}$ and $\kappa_{\rm{dr}}(d_0)\triangleq 1 + G_{\rm{sc}}g_{\text{r}}^{(0)} + N\frac{\pi}{4}\sqrt{\pi g_{\text{r}}^{(0)}} + NE_{\rm{S1}}(d_0)$ which depends on the locations of BS 0 and IRS 0. The dominant term in $\kappa_{\rm{dr}}(d_0)$ is $G_{\rm{sc}}g_{\text{r}}^{(0)}$ which scales in $O(N^2)$ or $O(d_0^{-\frac{\alpha}{2}})$ when $d_0$ is sufficiently small. For the large $d_0$, the IRS power gain is given by $G_{\rm{sc}}g_{\text{r}}^{(0)}\leq1$ and hence $\kappa_{\rm{dr}}(d_0)\approx1$. It reveals that when UE 0 is closer to IRS 0, IRS 0 provides more power gain.  \\ \indent
Then we derive the harvested signal power originated from all BSs $m\in\Lambda_{\rm{B}}\setminus\left\{0\right\}$ and reflected by IRSs $j\in\jmath$, which is given by
\vspace{-0.5em}
 \begin{equation}\label{equ40}
 	\small
 	\begin{aligned}
 		&\overline{S}_{\rm{id}}\triangleq\sum_{m\in\Lambda_{\rm{B}}\backslash\left\{0\right\}}\overline{S}_{\rm{id}}^{\left(m\right)}=
 		P_{\rm{t}}\left[\left(\sum_{m\in\Lambda_{\rm{B}}\backslash\left\{0\right\}}g_{{\rm{d}},m}\right) + N\sum_{j\in \jmath}\left(g_{\rm{r}}^{\left(j\right)}\sum_{m\in\Lambda_{\rm{B}}\backslash\left\{0\right\}}g_{{\rm{i}},m}^{\left(j\right)}\right)\right].
 	\end{aligned}
 \end{equation}
We consider the following approximation $g_{{\rm{i}},m}^{\left(j\right)}\approx g_{{\rm{d}},m}$ for $j\in\jmath$, and thus, we have $\overline{S}_{\rm{id}}^{\left(m\right)}\approx\upsilon_{\rm{S}}g_{{\rm{d}},m}$ and hence 
\vspace{-0.5em}
\begin{equation}\label{equ41}
	\small
	\overline{S}_{\rm{id}}\approx P_{\rm{t}}\upsilon_{\rm{S}}\sum_{m\in\Lambda_{\rm{B}}\backslash\left\{0\right\}}g_{{\rm{d}},m},
\end{equation}
where $\upsilon_{\rm{S}}\triangleq1+N\sum_{j\in \jmath}g_{\rm{r}}^{(j)}$ represents the relative power gain of reflecting or scattering paths originated from IRSs in $\jmath$ over the direct path. \\ \indent
To characterize the distribution of the signal power $S_{\rm{id}}$ originated directly from other BSs and reflected via all IRSs, we first derive the Laplace transform $\mathcal{L}_{S_{\rm{id}}\vert_{d_0,x_0}}\left(s\right)$ in Lemma 1, in which the instantaneous signal power $S_{\rm{id}}$ can be approximated by  
\vspace{-0.5em}
\begin{equation}\label{equ42}
	\small
	S_{\rm{id}}\approx P_{\rm{t}}\upsilon_{\rm{S}}\sum_{m\in\Lambda_{\rm{B}}\backslash\left\{0\right\}}g_{{\rm{d}},m}\omega_{m}.
	\vspace{-0.5em}
\end{equation} \indent
According to (\ref{equ42}), conditioned on $x_0$, $\mathbb{E}\{S_{\text{id}}\}$ is given by
\vspace{-0.5em}
\begin{equation}\label{sid}
	\small
	\begin{aligned}
		\mathbb{E}\{S_{\text{id}}\}|_{x_0}&\approx P_{\text{t}}\mathbb{E}\{\omega_{m}\}\mathbb{E}\{\upsilon_{\rm{S}}\}\mathbb{E}\left\{\sum_{m\in\Lambda_{\rm{B}}\backslash\left\{0\right\}}g_{{\rm{d}},m}\right\}\Bigg|_{x_0}\\& \overset{(a)}{=}P_{\text{t}}(1+NE_{S1}(0))2\pi\lambda_{\rm{B}}\int_{x=x_0}^{\infty}{g_d(x)x\text{d}x}=\frac{2\pi\lambda_{\rm{B}}\beta(1+NE_{S1}(0))P_{\text{t}}}{(\alpha-2)(x_0^2+H_{\text{B}}^2)^{\frac{\alpha}{2}-1}}
	\end{aligned}
\end{equation}
Note that (a) is due to the HPPP-distributed BSs locations and $\mathbb{E}\{\upsilon_S\}=NE_{S1}(0)$ based on (\ref{S1}).
$\mathbf{Lemma \;1.}$ \emph{Conditioned on $d_0$ and $x_0$, the Laplace transform of $S_{\rm{id}}$ is given by}
\vspace{-0.5em}
\begin{equation}\label{equ43}
	\small
	\mathcal{L}_{S_{\rm{id}}\vert_{d_0,x_0}}\left(s\right)\triangleq \mathbb{E}\left\{e^{-sS_{\rm{id}}}\right\}\vert_{d_0,x_0}\approx\exp\left(-2\pi\Lambda_{\rm{B}}U\left(sP_{\rm{t}}\upsilon_{\rm{S}}\right)\right),
\vspace{-0.5em}
\end{equation}
\emph{where the function $U\left(\cdot\right)$ is defined as}
\begin{equation}\label{equ44}
	\small
	\begin{aligned}
		U(x)&\triangleq \frac{\pi}{\alpha \sin(\frac{2\pi}{\alpha})}\left( \beta x\right)^{\frac{2}{\alpha}} - \frac{x_0^2 + H_{\rm{B}}^2}{2}\centerdot _2F_1\left( 1,\frac{2}{\alpha},1+\frac{2}{\alpha},-\frac{1}{g_{{\rm{d}},0}x}\right),
	\end{aligned}
\end{equation}
\emph{with $g_{{\rm{d}},0}=g_{\rm{d}}\left(x\right)\vert_{x=x_0}$ and $_2F_1$ being the Gauss hypergeometric function} \cite{ref32}. \\ \indent
$\emph{Proof}:$ The results can be proved by a minor modification of Proposition 4 in \cite{ref21}. We omit the full proof due to the space limitation.\qquad\qquad\qquad\qquad\qquad\qquad\qquad\qquad\qquad\qquad\qquad\qquad\quad\fbox{}  \\ \indent
Thus, the cumulative distribution function (CDF) of $S_{\rm{id}}\vert_{d_0,x_0}$ can be derived via the inverse Laplace transform of (\ref{equ43}), i.e.,
\begin{equation}\label{equ45}
	\small
	F_{S_{\rm{id}}\vert_{d_0,x_0}}\left(x\right)=\mathcal{L}^{-1}\left[\frac{1}{s}\mathcal{L}_{S_{\rm{id}}\vert_{d_0,x_0}}\left(s\right)\right]\left(x\right),
\end{equation}
which can be computed directly in standard computing software, such as Wolfram Mathematica. \\ \indent
We then study the UL signal power distribution in the UL stage. Similar to the charging stage, we still use Gamma distribution to approximate the conditional distribution $S_{\rm{UL}}$ which is given by
\begin{equation}\label{equ46}
	\small
	S_{\rm{UL}}|_{d_0,x_0} \overset{\text{approx.}}{\sim} \Gamma[k_{\rm{UL}},\theta_{\rm{UL}}].
\end{equation}
With the moment matching technique \cite{ref31}, we have
\begin{equation}\label{equ47}
	\small
	k_{\rm{UL}}\triangleq\frac{(\mathbb{E}\{S_{\rm{UL}}\}|_{d_0,x_0})^2}{\rm{var}\{S_{\rm{UL}}\}|_{d_0,x_0}},\quad \theta_{\rm{UL}}\triangleq\frac{\rm{var}\{S_{\rm{UL}}\}|_{d_0,x_0}}{\mathbb{E}\{S_{\rm{UL}}\}|_{d_0,x_0}}.
\end{equation}
\vspace{-0.5em}
To obtain $k_{\rm{UL}}$ and $\theta_{\rm{UL}}$, we derive the first and second moments of $S_{\rm{UL}}$ conditioned on $d_0$ and $x_0$.
 \begin{equation}\label{equ48}
 	\small
	\begin{aligned}
		&\mathbb{E}\left\{ S_{\rm{UL}}\right\}\vert_{d_0,x_0} \approx  P_0\cdot( \mathbb{E}\left\{ | b_1|^2\right\}|_{d_0,x_0} + \mathbb{E}\left\{ | b_2|^2\right\}|_{d_0,x_0}),
	\end{aligned}
\end{equation}
\begin{equation}\label{equ49}
	\small
	\begin{aligned}
		&\mathbb{E}\left\{ S_{\rm{UL}}^2\right\}|_{d_0,x_0} =P_0\cdot( \mathbb{E}\{ |b_1|^4\}|_{d_0,x_0} + \mathbb{E}\{ |b_2|^4\}|_{d_0,x_0} + 4\mathbb{E}\{ |b_1|^2\}|_{d_0,x_0}\mathbb{E}\{ |b_2|^2\}|_{d_0,x_0}).
	\end{aligned}
\end{equation}\indent
The first two moments of $S_{\rm{UL}}|_{d_0,x_0}$ are determined by the first two moments of $b_1$, $b_2$, $| b_1|^2$ and $| b_2|^2$. Following the similar approach to derive the first and second moment of $S_{\rm{dr}}\vert_{d_0,x_0}$ and the approximation of $r_{0,j}\approx x_0$,  $f_{r}^{\left(j\right)}\approx f_{{\rm{d}},0}$ for $j\in\jmath$, we can obtain the first two moments of  $b_1$, $b_2$, $| b_1|^2$, $| b_2|^2$ by replacing $a_1$, $a_2$, $g_{{\rm{d}},0}$ and $g_{\rm{r}}^{(0)}$ in (\ref{a1})--(\ref{a2^2}) with $b_1$, $b_2$, $f_{\text{d,0}}$ and $f_{\text{i,0}}^{(0)}$. \\ \indent
Finally, by integrating $\mathbb{E}\left\{S_{\rm{UL}}\right\}\vert_{d_0,x_0}$ over $d_0$ and $x_0$, we obtain  $\mathbb{E}\left\{S_{\rm{UL}}\right\}$ by
\vspace{-0.5em}
\begin{equation}\label{equ56}
	\small
	\mathbb{E}\left\{S_{\rm{UL}}\right\} = \int_{0}^{\infty}\int_{0}^{D}{\mathbb{E}\left\{S_{\rm{UL}}\right\}\vert_{d_0,x_0} f_{d_0}(d_0) f_{x_0}(x_0) {\rm d}d_0{\rm d}x_0},
\end{equation}
where $f_{d_0}(d_0)$ and $f_{x_0}(x_0)$, denote the probability density functions (PDFs) of $d_0$ and $x_0$,
\vspace{-0.5em}
\begin{equation}\label{distributiond}
	\small
	f_{d_0}(d)\triangleq 2\pi\lambda_{\rm{I}}de^{-\lambda_{\rm{I}}\pi d^2},\quad
	f_{x_0}(x)\triangleq 2\pi\lambda_{\rm{B}} xe^{-\lambda_{\rm{B}}\pi x^2}.
\end{equation}
\subsection{Interference Power Distribution}
We derive the interference power distribution in the UL stage in this subsection. Specifically, the interference is from active users transmitting information in the same time-frequency resource block.\\ \indent
Referring to (\ref{equ20}), the UE $k$-IRS $j$-BS 0 channel has been approximated by the CSCG distribution $\mathcal{CN}\left(0,Nf_{{\rm{i}},k}^{(j)}f_{\rm{r}}^{(j)}\right)$. Since both the direct interference channel and the cascaded UE $m$-IRS $j$-BS 0 channel follow the CSCG distribution, the composite interference channel $b_{{\rm{d}},k} + \sum_{j\in\jmath}{b_{{\rm{ir}},k}^{(j)}}$ from UE $k\in\Lambda_{\rm{u}}'\backslash\left\{0\right\}$  is the summation of independent CSCG RVs, which follows the CSCG distribution with zero mean and covariance $\mathbb{E}\left\{| b_{{\rm{d}},k}|^2\right\}+\sum_{j\in\jmath}{\mathbb{E}\left\{| b_{{\rm{ir}},k}^{(j)}|^2\right\}}$. Therefore, the composite interference power \( I_k = |b_{{\rm{d}},k} + \sum_{j\in\jmath}{b_{\rm{ir}}^{(j)}}|^2\) follows an exponential distribution with the expression given by
\vspace{-0.5em}
\begin{equation}\label{equ59}
	\small
	I_k\triangleq\overline{I}_k\xi_k = P_k\left( f_{{\rm{d}},k} + N\sum_{j\in\jmath}{f_{{\rm{i}},k}^{j}f_{\rm{r}}^{(j)}}\right)\xi_k,
\end{equation}
where $\overline{I}_k$ is the average interference power and $\xi_k\overset{\text{dist.}}{=} \xi \sim \exp(1)$. Therefore, given the locations of IoT devices and IRSs, the aggregated interference power from all the active IoT devices is the summation of exponential RVs which are independent while not identically distributed, following the Erlang distribution \cite{ref26}. Then, the mean conditional interference power can be derived by
\vspace{-0.5em}
\begin{equation}\label{equ60}
	\small
	\overline{I}\triangleq\sum_{k\in\Lambda_{\rm{u}}'\backslash\left\{0\right\}}{\overline{I}_k}= \sum_{k\in\Lambda_{\rm{u}}'\backslash\left\{0\right\}}P_k\left({f_{{\rm{d}},k} + N\sum_{j\in\jmath}{f_{{\rm{i}},k}^{(j)}f_{\rm{r}}^{(j)}}}\right).
\end{equation}
By using the approximation $r_{k,j}\approx y_k$, $f_{r}^{(j)}\approx f_{{\rm{d}},k}$ for $j\in\jmath$, we have $\overline{I}_k\approx \upsilon_{\rm{I}} f_{\text{d},k}$ and
\vspace{-0.5em}
\begin{equation}\label{equ61}
	\small
	\overline{I}\approx\upsilon_{\rm{I}}\sum_{k\in\Lambda_{\rm{u}}'\backslash\left\{0\right\}}{P_kf_{\text{d},k}},
\end{equation}
where $\upsilon_{\rm{I}}\triangleq 1 + N\sum_{j\in\jmath}{f_{{\rm{i}},0}^{(0)}}$ is the relative power gain of scattering paths with regards to all IRSs in $\jmath$ over that of the UE-BS direct path.  \\ \indent
To derive the spatial throughput later, we first obtain the interference power distribution conditioned on $d_0$ and $x_0$, which can be represented by the Laplace transform of the aggregated interference in Lemma 2. The aggregated interference can be expressed as
\vspace{-0.5em}
\begin{equation}\label{equ62}
	\small
	I\approx\upsilon_{\rm{I}}\sum_{k\in\Lambda_{\rm{u}}'\backslash\left\{0\right\}}{P_kf_{\text{d},k}\xi_k}.
\end{equation} \indent
$\mathbf{Lemma \;2.}$ \emph{The Laplace transform of interference power conditioned on $d_0$ and $x_0$ is given by}
\vspace{-0.5em}
\begin{equation}
	\small
	\mathcal{L}_{I|_{d_0,x_0}}\left(s\right) = \exp\left( -2\pi\lambda_{\rm{u}}'U_I\left(s\upsilon_{\rm{I}}\rho\right)\right),
\end{equation}
\emph{where the function $U_I(\cdot)$ is defined as}
\vspace{-0.5em}
\begin{equation}
	\small
	U_I(x)\triangleq \int_{0}^{\infty}\int_0^y {1- \frac{2\pi\lambda_{\rm{u}} D_ke^{-\lambda_{\rm{u}} \pi D_k^2}}{1+x\left(y^2+H_{\rm{B}}^2\right)^{-\alpha/2}\left(D_k^2+H_{\rm{B}}^2\right)^{\epsilon\alpha/2}}}{\rm d}D_k y{\rm d}y.
\end{equation}\indent
$\emph{Proof}:$ See Appendix \ref{A}. \qquad\qquad\qquad\qquad\qquad\qquad\qquad\qquad\qquad\qquad\qquad\qquad\qquad\qquad\fbox{} \\ \indent
Based on (\ref{equ62}) and  $\xi_k \overset{\text{dist.}}{=}\xi \sim \exp(1),\forall k$, the mean interference power is given by
\begin{equation}\label{equ65}
	\small
	\begin{aligned}
		\mathbb{E}\left\{I\right\} &\approx \mathbb{E}\left\{\xi\right\} \mathbb{E}\left\{\upsilon_{\rm{I}}\right\}\mathbb{E}\left\{\sum_{k\in\Lambda_{\rm{u}}'\backslash\left\{0\right\}}{f_{\text{d},k}}\right\} = \upsilon_{\rm{I}} E_{\rm{I}},
	\end{aligned}
\end{equation}
where $E_{\rm{I}}$  denotes the expectation of the summation of direct channel power from interfering UEs $k\in\Lambda_{\rm{u}}'\backslash\left\{0\right\}$ conditioned on $x_0$, which is given by
\begin{equation}\label{equ66}
	\small
	\begin{aligned}
			E_{\rm{I}} \overset{\text{(a)}}{=}2\pi\lambda_{\rm{u}}'\int_{y=0}^{\infty}\int_{D_k=0}^{y}2\pi\lambda_{\rm{u}} D_k e^{-\lambda_{\rm{u}}\pi D_k^2}\left(D_k^2+H_{\rm{B}}^2\right)^{\epsilon\alpha/2} \ \rho\left(y^2+H_{\rm{B}}^2\right)^{-\alpha/2}{\rm d}D_k y{\rm d}y.
	\end{aligned}
\end{equation}
Note that (a) is due to the PPP-distributed UEs locations, which helps calculate the aggregated interference over the 2D plane.  
\subsection{Coverage Probability and Spatial Throughput}
In this section, we derive the energy coverage probability and UL coverage probability. In separate stages based on the derived signal power distribution and interference power distribution, which leads to the overall coverage probability and spatial throughput eventually.
\subsubsection{Energy Coverage Probability}
Based on the derived channel power gain distribution in (\ref{equ34}), we derive the energy coverage probability in Theorem 1. \\ \indent
$\mathbf{Theorem \;1.}$ \emph{The probability that the energy harvested in the charging stage is higher than $E_{\rm{min}}$ is given by}
\vspace{-0.5em}
\begin{equation}\label{equ76}
	\small
	\bold{P}_{\rm{en}}=\int_{x_0=0}^{\infty}\int_{d_0=0}^{D}{\bold{P}_{\rm{en}}\vert_{d_0,x_0}f_{d_0}(d_0)f_{x_0}(x_0){\rm d}d_0{\rm d}x_0},
\end{equation}
\emph{where $\bold{P}_{\rm{en}}\vert_{d_0,x_0}$ is the conditional energy coverage probability given by}
\vspace{-0.5em}
\begin{equation}\label{lemma1}
	\small
	\begin{aligned}
			&\bold{P}_{\rm{en}}\vert_{d_0,x_0}\approx\mathbb{P}\left\{S_{\rm{dr}}>C(\tau)\right\}\vert_{d_0,x_0}+\mathbb{P}\left\{S_{\rm{id}}>C(\tau)\right\}\vert_{d_0,x_0} \\&+\mathbb{P}\left\{S_{\rm{dr}}+S_{\rm{id}}>C(\tau)\vert S_{\rm{dr}},S_{\rm{id}}<C(\tau)\right\}\vert_{d_0,x_0}\mathbb{P}\left\{S_{\rm{dr}}<C(\tau)\right\}\vert_{d_0,x_0}\mathbb{P}\left\{S_{\rm{id}}<C(\tau)\right\}\vert_{d_0,x_0}\\&-\mathbb{P}\left\{S_{\rm{dr}}>C(\tau)\right\}\vert_{d_0,x_0}\cdot\mathbb{P}\left\{S_{\rm{id}}>C(\tau)\right\}\vert_{d_0,x_0},
	\end{aligned}
\end{equation}
\emph{while for  $k_{\rm{dr}}$ is less than the given threshold $\widetilde{k}_{\rm{dr}}$, the conditional probabilities of \{$S_{\rm{dr}}>C(\tau)\}$, $\{S_{\rm{id}}>C(\tau)\}$ and $\{S_{\rm{dr}}+S_{\rm{id}}>C(\tau)\vert S_{\rm{dr}},S_{\rm{id}}<C(\tau)\}$ are, respectively, given by}
\vspace{-0.5em}
\begin{equation}\label{sc}
	\small
	\begin{aligned}
		\mathbb{P}\{S_{\rm{dr}}>C&(\tau)\}\vert_{d_0,x_0} \approx
		\sum_{i=0}^{k_{dr}-1}\frac{\left(-1\right)^i}{i!}\frac{\partial^i}{\partial s^i}\left[\mathcal{L}_{S_{\rm{id}}\vert_{d_0,x_0}}\left(\frac{s}{\theta_{\rm{dr}}}\right)\right]_{s=1},
	\end{aligned}
\end{equation}
\begin{equation}\label{ic}
	\small
	\begin{aligned}
			\mathbb{P}\left\{S_{\rm{id}}>C(\tau)\right\}\vert_{d_0,x_0} \approx 1-\mathcal{L}^{-1}\left[\frac{1}{s}\mathcal{L}_{S_{\rm{id}}\vert_{d_0,x_0}}\left(s\right)\right]\left(z\right),
	\end{aligned}
\end{equation}
\begin{equation}\label{sic}
	\small
	\begin{aligned}
		\mathbb{P}\left\{S_{\rm{dr}}+S_{\rm{id}}>C(\tau)\vert \right.&\left. S_{\rm{dr}},S_{\rm{id}}<C(\tau)\right\}\vert_{d_0,x_0}\approx \\& \sum_{i=0}^{k_{\rm{dr}}-1}{\frac{\left( -1 \right) ^i}{i!}\frac{\partial i}{\partial s^i}\left[ \exp\left( -\frac{sC(\tau)}{\theta_{\rm{dr}}}+2\pi \lambda _BU\left( \frac{s\upsilon_S}{\theta _{\rm{dr}}} \right) \right)\right] _{s=1}}.
	\end{aligned}
\end{equation} \indent
$\emph{Proof}:$ See Appendix \ref{B}.\qquad\qquad\qquad\qquad\qquad\qquad\qquad\quad\qquad\qquad\qquad\qquad\qquad\qquad\qquad\fbox{} \\ \indent
\textcolor{blue}{\textbf{Remark} 1: \emph{It is worth noting that there exists correlation between the two events $\{S_{\rm{dr}}>C(\tau)\}$ and $\{S_{\rm{id}}>C(\tau)\}$, which is extremely hard to derive. Thus, to simplify the analysis, we ignore the correlation and assume that the two events are independent. We will show that the approximation is acceptable in Fig. \ref{energyn1000} in the numerical results part.} \qquad\qquad\qquad \qquad\qquad\qquad\qquad\qquad\qquad\qquad\fbox{}}
\subsubsection{UL Coverage Probability}
Due to the correlation between the locations of BSs and scheduled IoT devices, it's challenging to derive the exact UL analysis \cite{ref13}. The authors in \cite{ref13} proposed an approximation method by assuming that the locations of devices follow a PPP and handling the dependence between the typical link length and interferer link length. With such an approximation, we derive the UL coverage probability in Theorem 2.  \\ \indent
$\mathbf{Theorem \;2.}$ \emph{The probability that the received SINR of the typical UE 0 is greater than a given threshold $\overline{\gamma}$ is given by}
\vspace{-0.5em}
\begin{equation}\label{equ80}
	\small
	\begin{aligned}
		\bold{P}_{\rm{UL}}=\int_{x_0=0}^{\infty}\int_{d_0=0}^{D}{\bold{P}_{\rm{UL}}\vert_{d_0,x_0}f_{d_0}(d_0)f_{x_0}(x_0){\rm d}d_0{\rm d}x_0},
	\end{aligned}
\end{equation} 
\emph{where $\bold{P}_{\rm{UL}}\vert_{d_0,x_0}$ is the conditional coverage probability given by}
\vspace{-0.5em}
\begin{equation}\label{lemma2}
	\small
	\begin{aligned}
		  &\bold{P}_{\rm{UL}}\vert_{d_0,x_0}\approx \sum_{i=0}^{k_{\rm{UL}}-1}{\frac{\left(-1\right)^i}{i!}\frac{\partial^i}{\partial s^i}\left[-\frac{s\overline{\gamma}W}{\theta_{\rm{UL}}}-2\pi\lambda_{\rm{u}}'U_I\left(\frac{s\overline{\gamma}\upsilon_{\rm{I}}\rho}{\theta_{\rm{UL}}}\right)\right]_{s=1}}  .
	\end{aligned}
\end{equation} \indent
$\emph{Proof}:$ See Appendix \ref{C}. \qquad\qquad\qquad\qquad\qquad\qquad\qquad\qquad\qquad\qquad\qquad\qquad\qquad\qquad\quad\fbox{}\\ \indent 
\textcolor{blue}{\textbf{Remark 2:} \emph{The results derived in (\ref{equ80}) degenerates into the traditional regularly powered IRS-assisted UL transmission case without EH, by setting $\bold{P}_{\text{en}}=1$ and $\lambda_{\rm{u}}^{'} =\lambda_{\rm{B}}$}. \qquad\qquad\qquad\qquad\quad\fbox{}} \\ \indent
In both Theorem 1 and Theorem 2, we need to discuss the scale parameters $k_{dr}$ and $k_{UL}$ of Gamma distribution.  For non-integer $k_{dr}(k_{UL})$, its upper and lower bounds are utilized to approximate the conditional energy (UL) coverage probability. 
Take Theorem 1 as an example, we approximate $\bold{P}_{\rm{en}}\vert_{d_0,x_0}$ by the following linear combination \cite{ref21},
\vspace{-0.5em}
\begin{equation}\label{equ74}
	\small
	\bold{P}_{\rm{en}}\vert_{d_0,x_0} = \varOmega \bold{P}_{\rm{en}}\vert_{d_0,x_0, \lfloor k \rfloor} + \left(1-\varOmega\right)\bold{P}_{\rm{en}}\vert_{d_0,x_0,\lceil k\rceil},
\end{equation}
where $\lceil \cdot \rceil$ and $\lfloor \cdot \rfloor$ are the ceiling and floor functions, and $\varOmega$ is the weight between the ceiling and is the weight functions given by
\vspace{-1em}
\begin{equation}\label{equ75}
	\small
	\varOmega \triangleq \frac{\zeta\left( \lceil k \rceil -k \right)}{\zeta\left( \lceil k \rceil -k \right) +\left( k-\lfloor k \rfloor \right)},	
\end{equation}
where $\zeta>0$ is a parameter to illustrate the relationship between $\bold{P}_{\rm{en}}\vert_{d_0,x_0}$ and $k$. The weights $\varOmega$ and $\left(1-\varOmega\right)$ are set as \( \frac{\varOmega}{1-\varOmega}=M\frac{ \lceil k \rceil -k}{ k-\lfloor k \rfloor }\). \\ \indent
For the large value of $k$, we approximate the Gamma distribution to the normal distribution with $\mu \triangleq \mathbb{E}\left\{S\right\}\vert_{d_0,x_0}$ as the mean value, and $\sqrt{\rm{var}\left\{S\right\}\vert_{d_0,x_0}}$ as the standard deviation. Since the former is much larger than the latter, $S$ can be approximated by $\mu$.\\ \indent
According to Theorem 1 and Theorem 2, we derive the overall coverage probability in the next subsection.
\vspace{-0.5em}
\subsubsection{Overall Coverage Probability and Spatial Throughput}
With the derived energy coverage probability and UL coverage probability, we obtain the overall coverage probability in this subsection.  \\ \indent
$\mathbf{Corollary \;1.}$ \emph{The overall coverage probability $P_{\rm{cov}}$ of the typical UE 0 is }
\vspace{-1em}
\begin{equation}\label{equ81}
	\small
	\begin{aligned}
			\bold{P}_{\rm{cov}}&=\bold{P}_{\rm{en}}\times\bold{P}_{\rm{UL}} ,
	\end{aligned}
\vspace{-0.5em}
\end{equation} 
\emph{where $\bold{P}_{\rm{en}}$ and $\bold{P}_{\rm{UL}}$ are given by (\ref{equ76}) and (\ref{equ80}), respectively.} \\ \indent
$\emph{Proof}$: By definition, the overall coverage probability can be derived by  multiplying the energy coverage probability by the UL coverage probability.\qquad\qquad\qquad\qquad\qquad\qquad\qquad\qquad\qquad\quad\fbox{}  \\ \indent
$\mathbf{Remark\; 3.}$ \emph{Note that the proposed analytical framework can be used to characterize the regularly powered IRS-assisted UL IoT network performance by specializing some system parameters. To be specific, the parameters that leads to the result $\bold{P}_{\rm{en}} \to1$, for instance, a large BS density $\lambda_{\rm{B}}$, a large number of reflecting element $N$, or a sufficiently large $\tau$ result in the regularly powered IRS-assisted UL IoT network. In this case, the coverage probability is equivalent to $\bold{P}_{\rm{UL}}$ in (\ref{equ80}) by setting $\lambda_{\rm{u}} = \lambda_{\rm{B}}$.} \qquad\qquad\qquad\qquad\qquad\qquad\qquad\qquad\qquad\qquad\qquad\qquad\qquad\qquad\qquad\qquad\quad\fbox{} \\ \indent
By definition, the spatial throughput $\nu$ in (\ref{equ82}) can be derived by substituting $\bold{P}_{\rm{en}}$ and $\bold{P}_{\rm{cov}}$ in (\ref{equ76}) and (\ref{equ81}), respectively. The complete expression of spatial throughput is omitted to keep concise. \\ \indent
\subsubsection{Power Efficiency}
\textcolor{blue}{The EHE in charging stage and EE in UL stage can be calculated by substituting the expressions of $\mathbb{E}\{E_{\text{h}}\}$ and $\mathbb{E}\{P_0\}$ into (\ref{ehef}) and (\ref{eef}), which are given by,
\vspace{-0.5em}
\begin{equation}
	\small
	\mathbb{E}\{E_{\text{h}}\} =\int_{0}^{\infty}\int_{0}^{D} \tau T\eta \left(\mathbb{E}\{S_{\rm{dr}}\}|_{d_0,x_0}+ \mathbb{E}\{S_{\rm{id}}\}|_{x_0}\right)f_{d_0}(d_0)f_{x_0}(x_0)\text{d}d_0\text{d}x_0,
\end{equation}
\begin{equation}
	\small
	\mathbb{E}\{P_0\}=\int_{0}^{\infty}\rho (x_0^2+H_{\text{B}}^2)^{\epsilon\alpha/2}f_{x_0}(x_0)\text{d}x_0,
\end{equation}
where $f_{x_0}(x_0)$ $(f_{d_0}(d_0))$, $\mathbb{E}\{S_{\text{dr}}\}|_{d_0,x_0}$ and $\mathbb{E}\{S_{\text{id}}\}|_{x_0}$ are given by (\ref{distributiond}), (\ref{equ36}) and (\ref{sid}), respectively.
} \\ \indent
\textcolor{blue}{\textbf{Remark 4.} \emph{The following technical challenges should be addressed when modifying the proposed analytical framework to adapt to the multiple-antenna BS case. Firstly, both active beamforming precoder/receiver at BSs and passive beamforming design of IRSs should be considered, and jointly optimized to maximize the network performance in charging stage and UL stage. Secondly, the signal power distributions in both charging stage and UL stage, and the interference distribution in UL stage are dependent on the joint beamformer design, which is unknown for the multiple-antenna BS case in the large-scale network scenario.}\qquad\qquad\qquad\qquad\qquad\qquad\qquad\qquad\qquad\qquad\qquad\qquad\qquad\quad\fbox{}} 
\begin{figure}[htbp]
	\vspace{-1em}
	\centering
	\begin{minipage}[t]{0.49\textwidth}
		\includegraphics[scale=0.6]{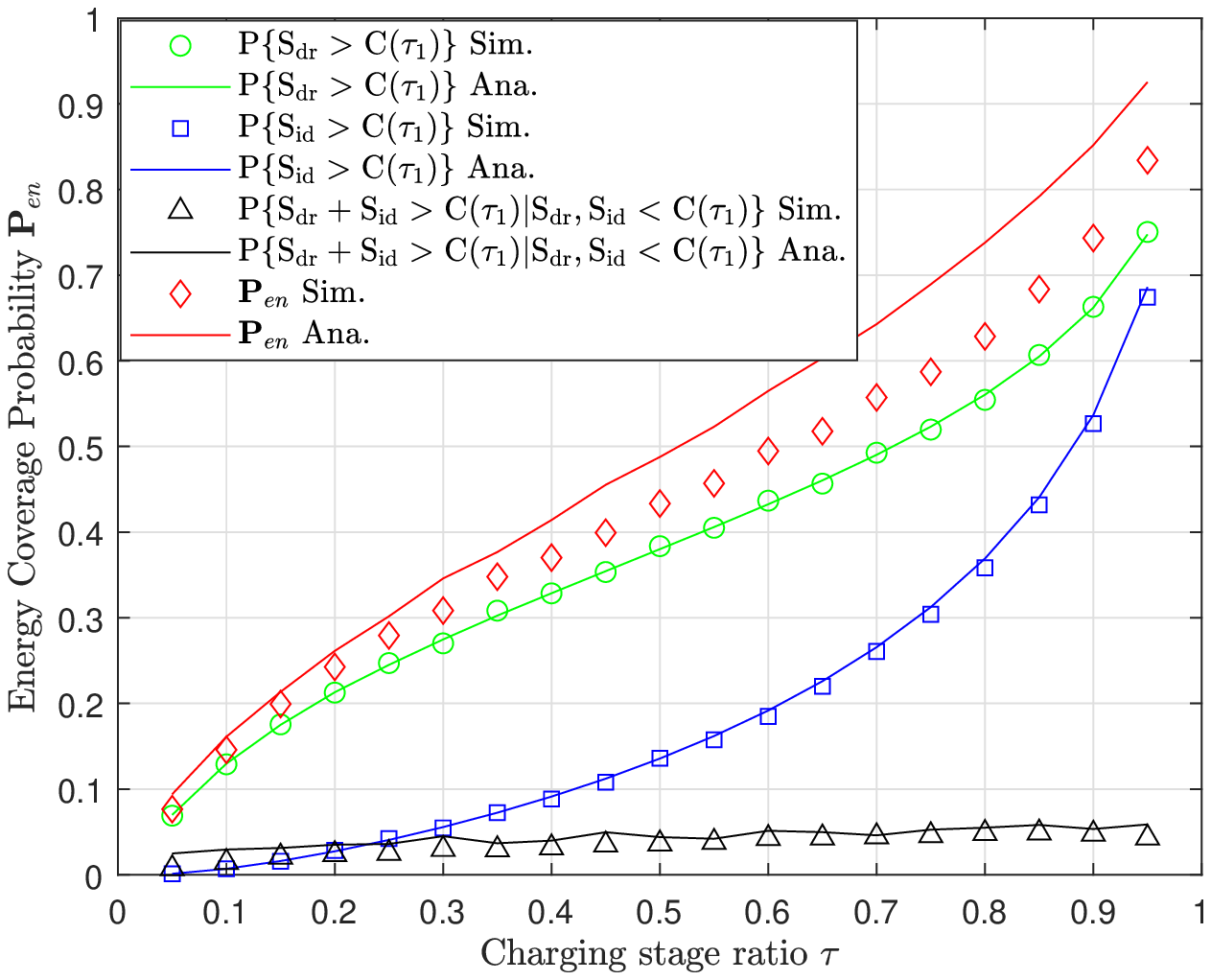}
		\caption{Energy coverage probability vs. $\tau$.}
		\label{energyn1000}
	\end{minipage}
	\begin{minipage}[t]{0.5\textwidth}
		\includegraphics[scale=0.6]{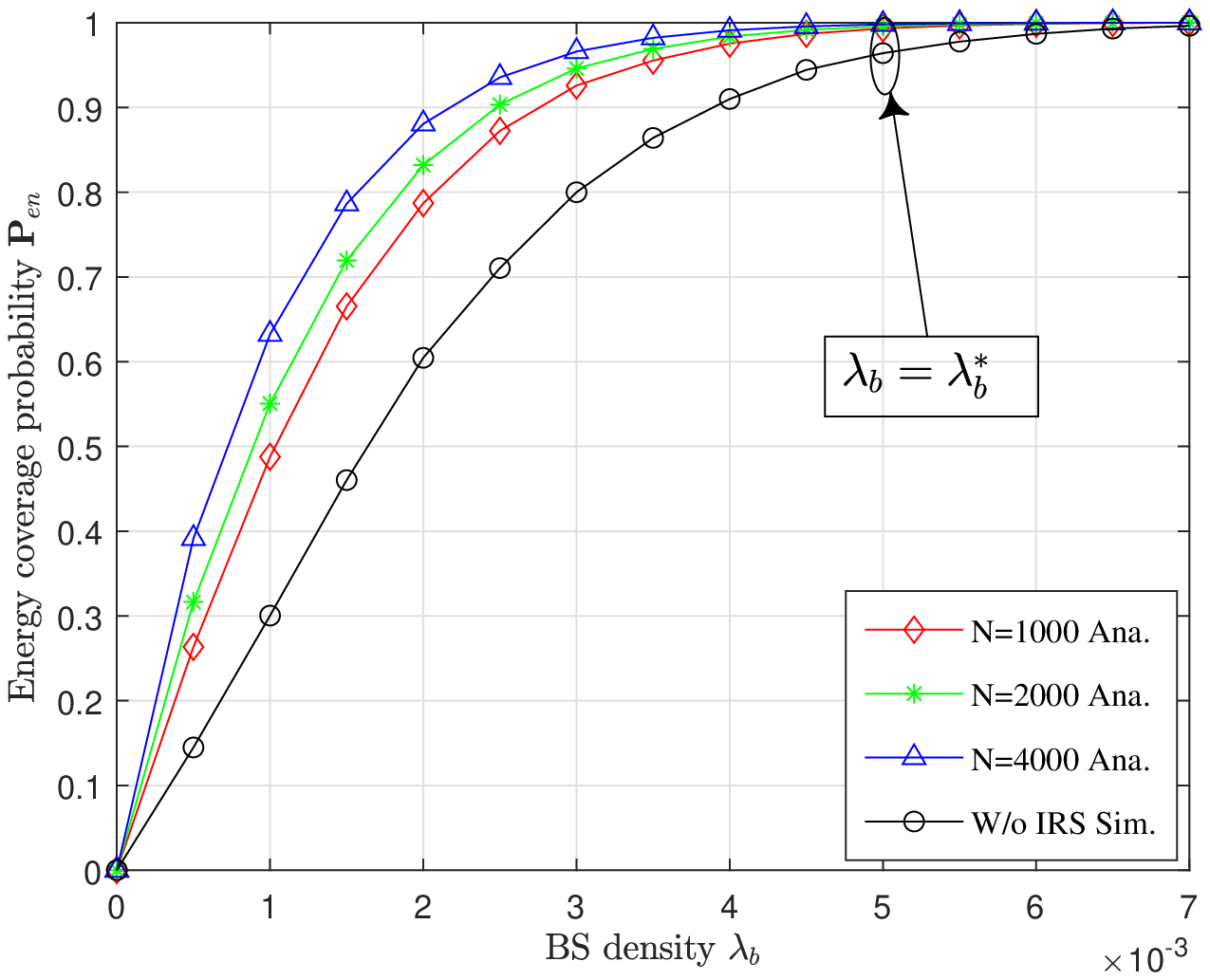}
		\caption{Energy coverage probability as a function of SINR threshold for different IRS reflecting element numbers $N$.}
		\label{energycontrast}
	\end{minipage}
\end{figure}
\vspace{-1em}
\section{NUMERICAL RESULTS}
In this section, the analytical (Ana.) framework is verified via extensive simulations (Sim.) by evaluating the network performance in terms of energy coverage probability, UL coverage probability, overall coverage probability, network spatial throughput \textcolor{blue}{and power efficiency}, respectively. \textcolor{blue}{To show the superiority of IRSs in enhancing the coverage probability, we choose the traditional RF-powered IoT network considered in \cite{joint} as a benchmark. Compared to the IRS-assisted RF-powered IoT network, the only difference is that there are no IRSs existing in the the traditional RF-powered IoT network, denoted by "W/o IRS" in the simulations.} Unless otherwise specified, we use the default values of the system parameters as follows: $H_{\rm{B}}=10\;\text{m}$, $H_{\rm{I}}=1\;\text{m}$, $\lambda_{\rm{B}}=10^{-3}\; \text{m}^2$, $\lambda_{\rm{I}}=10^{-2}\; \rm{m}^2$, $\alpha=4$, $P_{\rm{t}}=46\; \text{dBm}$, $N=1000$, $f_c=2\; \text{GHz}$, $\sigma^2=-147\;\text{dBm}$, $D=25\;\text{m}$, $\rho=-78.5\; \text{dBm}$, $\eta=0.5$, $\epsilon=0.8$, $T=0.01\;\text{s}$, $\tau=0.5$, \textcolor{blue}{$P_{\text{cb}}=2.5$ W, $P_{\text{cu}}=20$ mW \cite{r31} and $\eta_{\text{b}}=\eta_{\text{u}}=0.2$ \cite{r30}}.
\vspace{-1.5em}
\subsection{Energy Coverage Probability}
In this subsection, we characterize the impact of IRSs in the charging stage by setting $\zeta=0.5$ in (\ref{equ75}). Fig. \ref{energyn1000} depicts the energy coverage probability $\bold{P}_{\rm{en}}$ given in (\ref{equ76}) in Theorem 1 along with the probabilities of events $\{S_{\rm{dr}}>C(\tau)\}$, $\{S_{\rm{id}}>C(\tau)\}$, and $\{S_{\rm{dr}}+S_{\rm{id}}>C(\tau)|S_{\rm{dr}}<C(\tau),S_{\rm{id}}<C(\tau)\}$, respectively. We observe a very perfect match between the theoretical analysis and the simulation results except for $\bold{P}_{\rm{en}}$. The reason for this is that we ignore the correlation between the events $\{S_{\rm{dr}}>C(\tau)\}$ and $\{S_{\rm{id}}>C(\tau)\}$ to simplify the analysis, and the independent approximation leads to an acceptable gap. In Fig. \ref{energyn1000}, we observe that $\bold{P}_{\rm{en}}$ increases with the time slot ratio $\tau$. This is intuitive since a larger $\tau$ provides more opportunities for an IoT device to harvest energy, leading to a higher $\bold{P}_{\rm{en}}$. \\ \indent 
In Fig. \ref{energycontrast}, we quantify the gains achieved by IRS beamforming in the charging stage by varying the IRS reflecting element number $N$. We observe that increasing BS density $\lambda_b$ or IRS reflecting element number $N$ leads to a higher $\bold{P}_{\rm{en}}$. This can be explained by the fact that a denser BS deployment or a larger IRS enhances the harvested RF power. Compared to the conventional network scenario without IRS deployment, IRS-assisted EH scheme with passive beamforming results in a more than 30$\%$ enhancement in $\bold{P}_{\rm{en}}$ for $N=4000$ with BS density $\lambda_b$ = $10^{-3}$ $\rm{m}^{-2}$. Moreover, when $\lambda_b$ achieves $\lambda_b^{*}$, i.e., $\lambda_b^*=5\times10^{-3}$  $\rm{m}^{-2}$ for the given system parameters, the energy coverage condition $\{E_{\rm{h}} \geq E_{\rm{min}}\}$ satisfies with an extremely high probability, resulting in $\bold{P}_{\rm{en}} \to 1$.
\begin{figure}[htbp]
	\vspace{-1em}
	\centering
	\begin{minipage}[t]{0.49\textwidth}
		\includegraphics[scale=0.6]{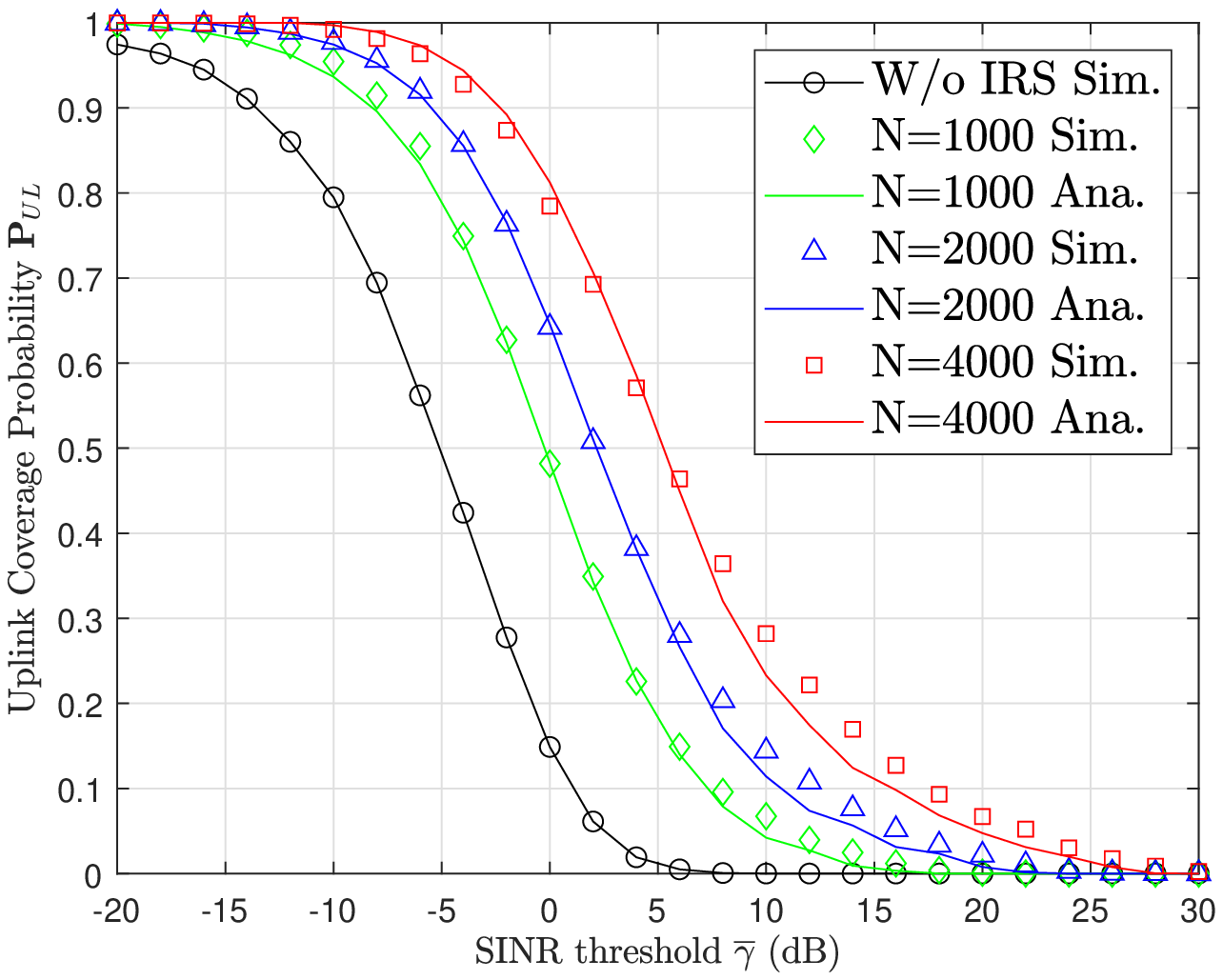}
		\caption{UL Coverage Probability as a function of SINR threshold for different IRS reflecting element numbers $N$.}
		\label{uplinkSINR}
	\end{minipage}
	\begin{minipage}[t]{0.5\textwidth}
		\includegraphics[scale=0.6]{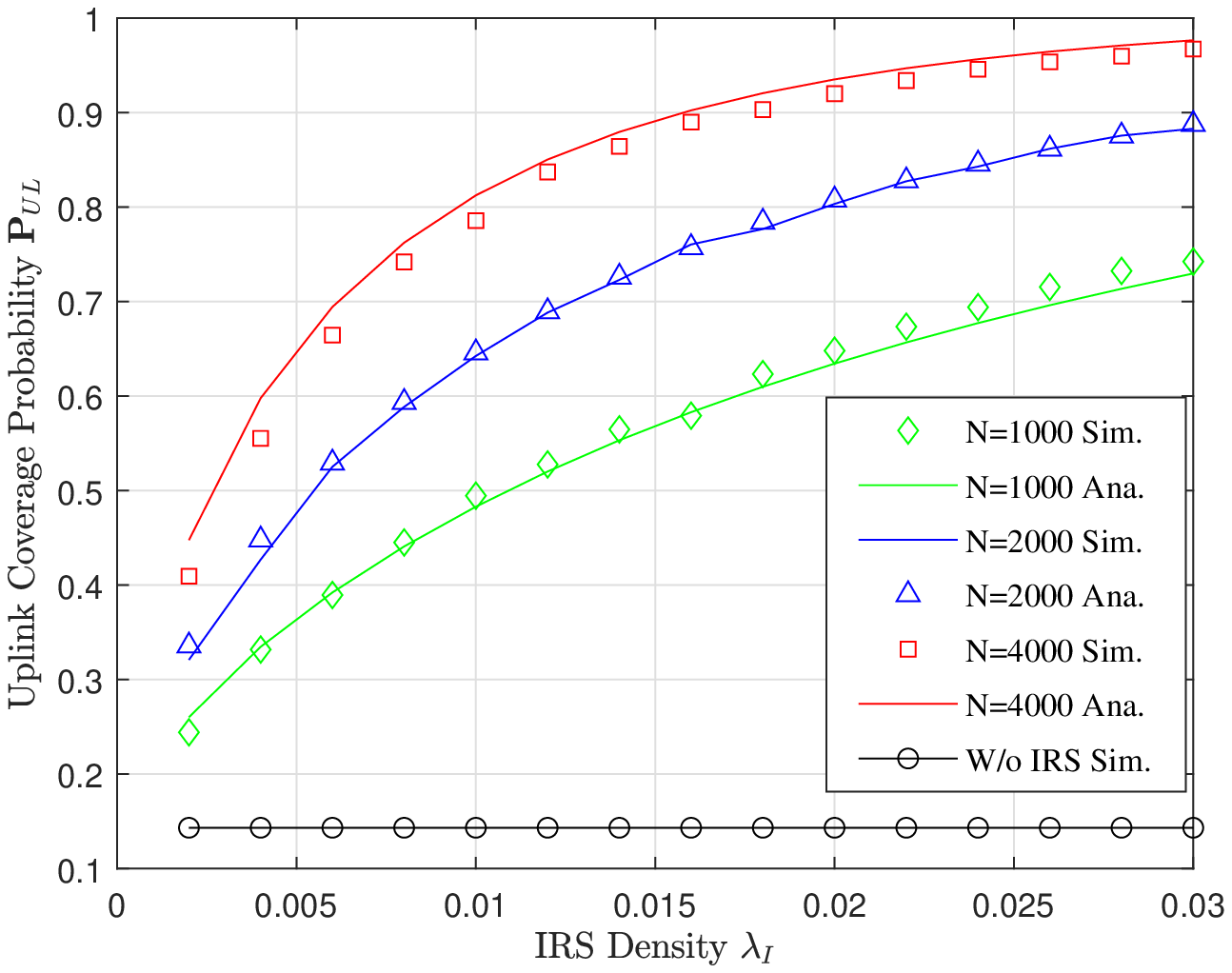}
		\caption{UL Coverage Probability as a function of IRS density $\Lambda_{\rm{I}}$ for different IRS reflecting element numbers $N$.}
		\label{lamdar}
	\end{minipage}
\end{figure}
\vspace{-1.5em}
\subsection{UL Coverage Probability} 
In this subsection, we focus on the performance analysis of the UL transmission mode by setting $\zeta=1$ in (\ref{equ75}). In Fig. \ref{uplinkSINR}, the UL coverage probability is evaluated by varying SINR threshold $\overline{\gamma}$ and IRS reflecting element number $N$, with the conventional UL coverage probability without IRSs being given as a benchmark. We observe that $\bold{P}_{\rm{UL}}$ reduces with the increasing SINR threshold $\overline{\gamma}$, and grows with the element number $N$. This is because the passive beamforming gain provided by IRSs increases with $N$ in the order of $O(N^2)$ which enlarges the UL received signal power and hence the UL coverage probability. Compared to the benchmark, the coverage enhancement achieved by IRS passive beamforming can be as large as 0.65 with $N=4000$ for $\gamma = 0$ dB.\\ \indent
 In Fig. \ref{lamdar}, we evaluate the UL coverage probability $\bold{P}_{\rm{UL}}$ as a function of IRS density $\lambda_{\rm{I}}$ and IRS reflecting element number $N$. From Fig. \ref{lamdar}, we find that $\bold{P}_{\rm{UL}}$ grows with the increasing $\lambda_{\rm{I}}$ and $N$. This can be explained by the fact that $\lambda_{\rm{I}}$ is large enough to achieve a full energy coverage, i.e., $\bold{P}_{\text{en}}=1$, and passive beamforming gain provided by IRS dominates the growing interference contributed by IRS random scattering, resulting in a higher $\bold{P}_{\rm{UL}}$. \\ \indent
 The mean signal power $\mathbb{E}\left\{ S_{\rm{UL}}\right\}$ and interference $\mathbb{E}\{\rm{I}\}$ derived in (\ref{equ56}) and (\ref{equ65}) by varying $\lambda_{\rm{I}}$ and $N$ are illustrated in Fig. \ref{signalpower}. We observe that $\mathbb{E}\left\{ S_{\rm{UL}}\right\}$ increases with the growing $\lambda_{\rm{I}}$, while $\mathbb{E}\{\rm{I}\}$ remains almost unchanged. This is consistent with the observations in Fig. \ref{energycontrast} and Fig. \ref{uplinkSINR}.
 \vspace{-1.5em}
\subsection{Overall Coverage Probability and Network Spatial Throughput}
 \begin{figure}[h]
	\vspace{-1em}
	\centering
	\begin{minipage}[t]{0.49\textwidth}
		\includegraphics[scale=0.6]{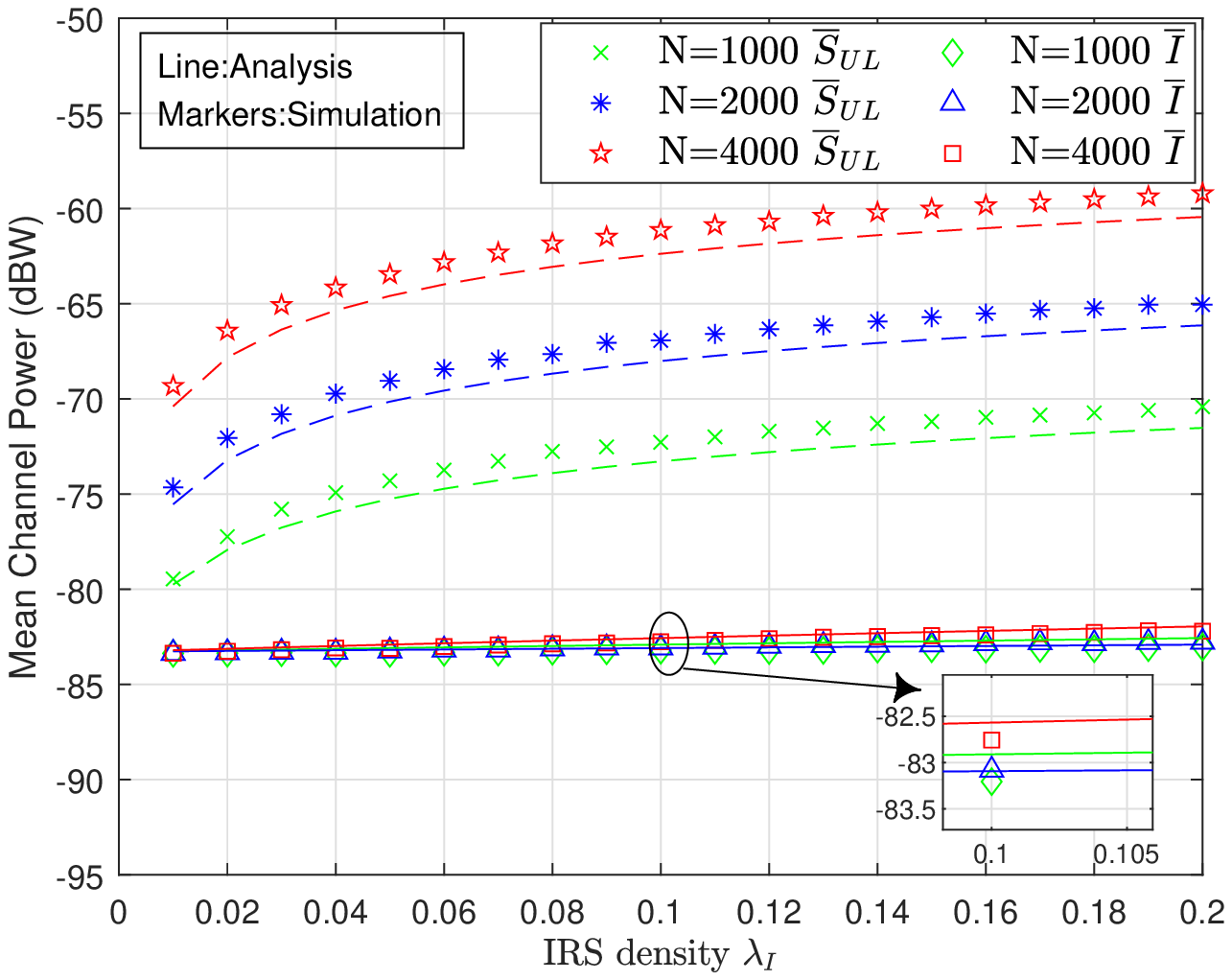}
		\caption{Signal power and interference power as a function of IRS density $\lambda_{\rm{I}}$ for different IRS reflecting element numbers $N$.}
		\label{signalpower}
	\end{minipage}
	\begin{minipage}[t]{0.5\textwidth}
		\includegraphics[scale=0.6]{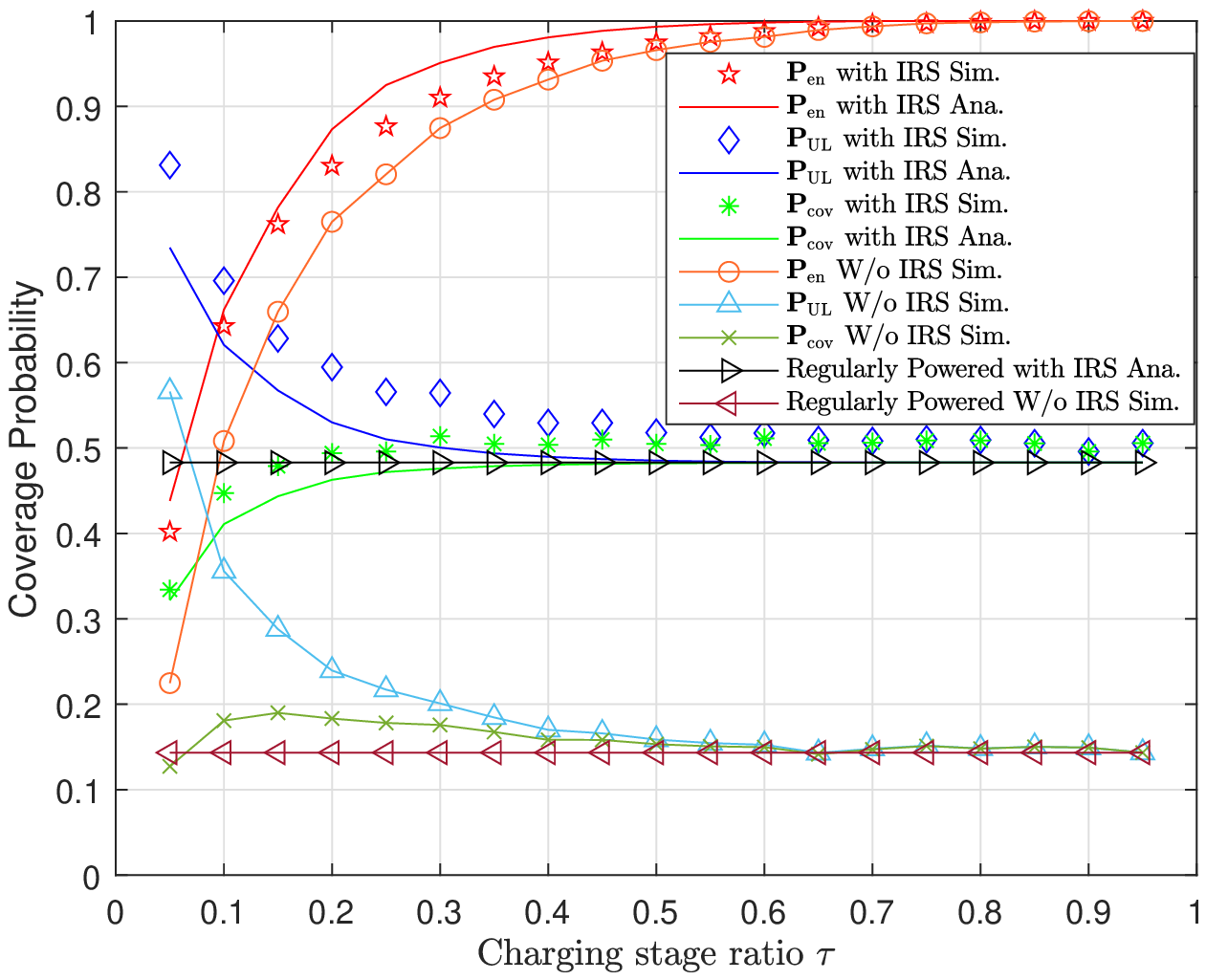}
		\caption{Coverage probability vs. charging stage ratio $\tau$.}
		\label{wholeprocess}
	\end{minipage}
\end{figure}
\begin{figure}
	\vspace{-2em}
	\centering
	\begin{minipage}[t]{0.49\textwidth}
		\includegraphics[scale=0.6]{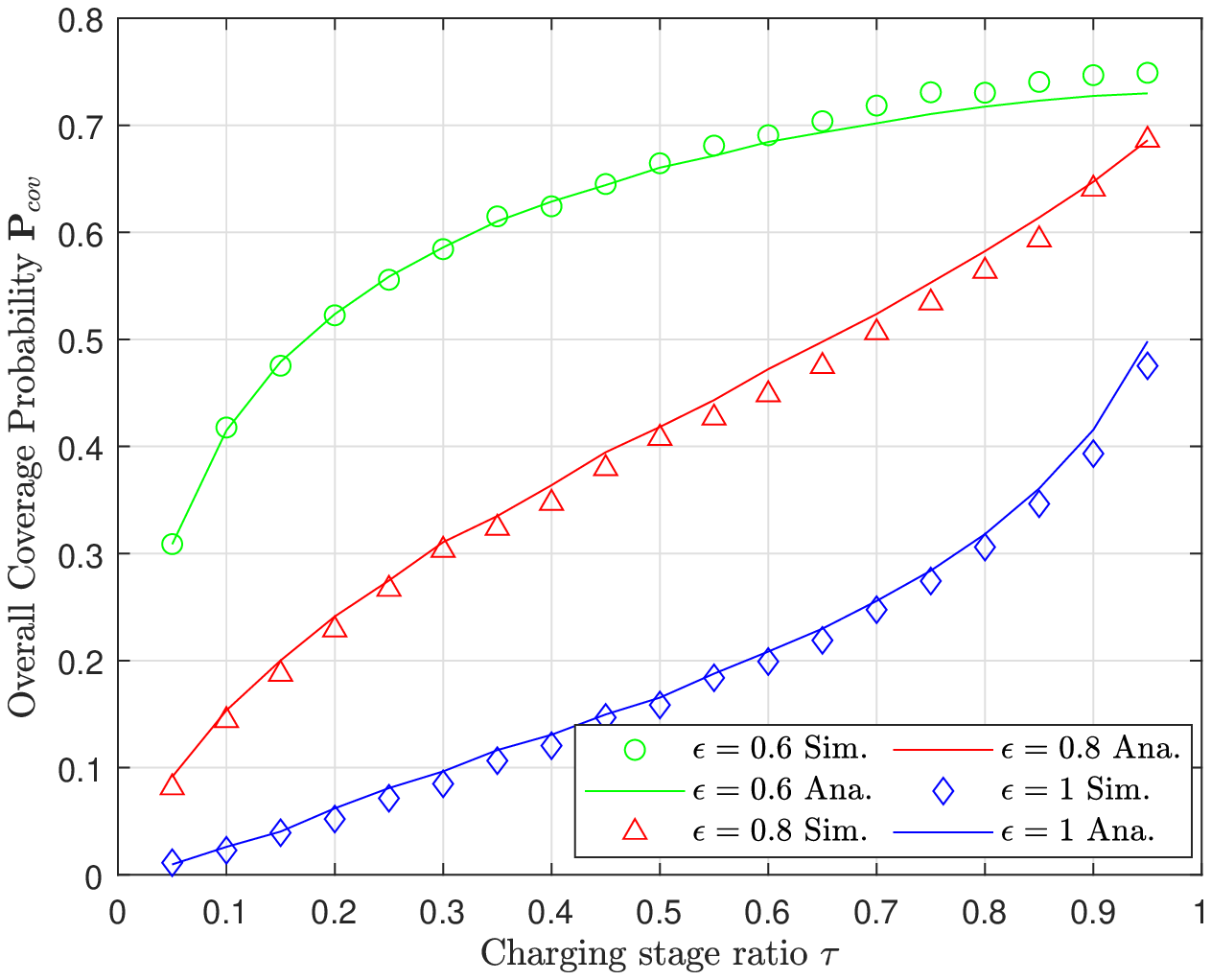}
		\caption{The overall coverage probability vs. charging stage ratio $\tau$ for different power control factors 	$\epsilon$.}
		\label{totalepsilon}
	\end{minipage}
	\begin{minipage}[t]{0.5\textwidth}
		\includegraphics[scale=0.6]{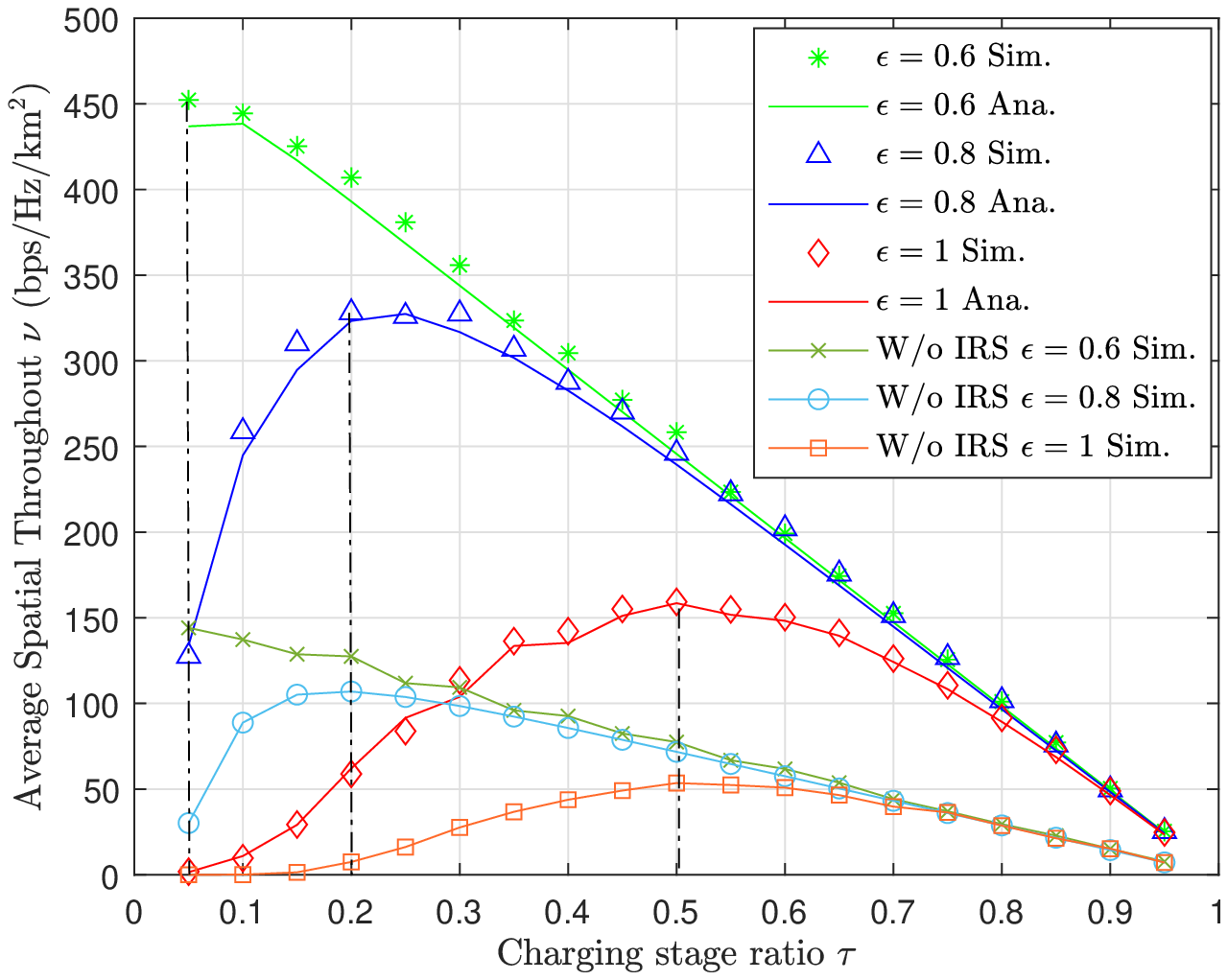}
		\caption{Average spatial throughput as a function of charging stage ratio $\tau$ for different power control factors $\epsilon$.}
		\label{throughput}
	\end{minipage}
\end{figure}

In this subsection, we focus on the overall coverage probability and network spatial throughput. In Fig. \ref{wholeprocess}, we set BS density $\lambda_B = 5\times10^{-3}$ $\rm{m}^{-2}$ and evaluate $\bold{P}_{\rm{en}}$, $\bold{P}_{\rm{UL}}$ and $\bold{P}_{\rm{cov}}$ with and without IRSs as a function of charging stage ratio $\tau$. To show the availability of the proposed IRS-assisted RF EH scheme, we plot the coverage probability with regularly powered IoT devices as the benchmark. In this case, all IoT devices are full of energy which is equivalent to $\bold{P}_{\rm{en}}=1$. We observe that when $\tau$ reaches a certain value, $\bold{P}_{\rm{en}}$ approaches to 1 and hence, $\bold{P}_{\rm{UL}}$ approaches to $\bold{P}_{\rm{cov}}$. This is because although a growing $\tau$ enlarges the density of interferer which declines $\bold{P}_{\rm{UL}}$, the improvement in $\bold{P}_{\rm{en}}$ dominates the reduction in $\bold{P}_{\rm{UL}}$, leading to the growth in $\bold{P}_{\rm{cov}}$. By comparing with the traditional network without IRS deployment, we observe the significant superiority of IRS passive beamforming in both $\bold{P}_{\rm{en}}$ and $\bold{P}_{\rm{UL}}$. By comparing with the regularly powered benchmark, we observe that when $\tau$ reaches a certain value, i.e., $\tau = 0.4$ for the IRS case, the RF powered IoT network achieves nearly the same overall coverage probability as that of the regularly powered IoT network. What's more, even for the regularly powered IoT network, IRSs passive beamforming brings huge enhancement in coverage, more than 0.3 with the given parameters. The comprehensiveness of the proposed analytical framework lies in that it degenerates into the IRS-assisted regularly-power IoT network when $\bold{P}_{\rm{en}}=1$ is satisfied. \\ \indent
 Fig. \ref{totalepsilon} depicts the overall coverage probability $\bold{P}_{\rm{cov}}$ by varying charging stage ratio $\tau$ for different power control factors $\epsilon$. We observe that a larger $\epsilon$ results in a lower $\bold{P}_{\rm{cov}}$. Note that $E_{\rm{min}} = (1-\tau)T\rho (x_0^2+H_{\text{B}}^2)^{\epsilon \alpha/2}$, a larger $\epsilon$ leads to an increasing $E_{\rm{min}}$, which results in the decline of $\bold{P}_{\rm{en}}$, leading to a lower $\bold{P}_{\rm{cov}}$.  \\ \indent
In Fig. \ref{throughput}, we evaluate the average spatial throughput $\nu$ as a function of charging stage ratio $\tau$ and power control factor $\epsilon$. We observe that there exists an optimal $\tau$ to maximize the average spatial throughput $\nu$. A larger $\epsilon$ requires a higher proportion $\tau$ to achieve the largest spatial throughput. Referring to (78), the average spatial throughput is related to the product of $(1-\tau)$, $\bold{P}_{\rm{en}}$ and $\bold{P}_{\rm{cov}}$. In addition, a tradeoff exists between $\tau$ and $\bold{P}_{\rm{en}}$. When $\epsilon$ = 0.6, $E_{\rm{min}}$ is easy to satisfy, and the spatial throughput is dominated by $\tau \bold{P}_{\rm{cov}}$. On the other hand, as $\epsilon$ further enlarges, $E_{\rm{min}}$ grows exponentially and the spatial throughput is limited by $\bold{P}_{\rm{en}}$. In this case, a larger $\tau$ is required to achieve a higher spatial throughput. As $\bold{P}_{\rm{en}}$ grows to a certain value, $\tau$ dominates the spatial throughput, resulting in the decline of spatial throughput. Compared to the case without IRSs, we observe that deploying IRSs can greatly enhance the achievable network spatial throughput, while does not change the trend of $\nu$ as a function of the charging stage ratio $\tau$, and hence the the optimal $\tau$ that maximizes $\nu$. This is because with IRS passive beamforming, both the harvested signal power in the charging stage and the desired signal power in the UL stage are shown to scale with $N$ in the same order of $O(N^2)$, while only slightly increases the interference power, leading to the same level of enhancements in $\bold{P}_{\rm{en}}$ and $\bold{P}_{\rm{UL}}$.
\vspace{-1.5em}
\subsection{Power Efficiency}
\begin{figure}[htbp]
	\vspace{-2em}
	\centering
	\begin{minipage}[t]{0.49\textwidth}
		\includegraphics[scale=0.52]{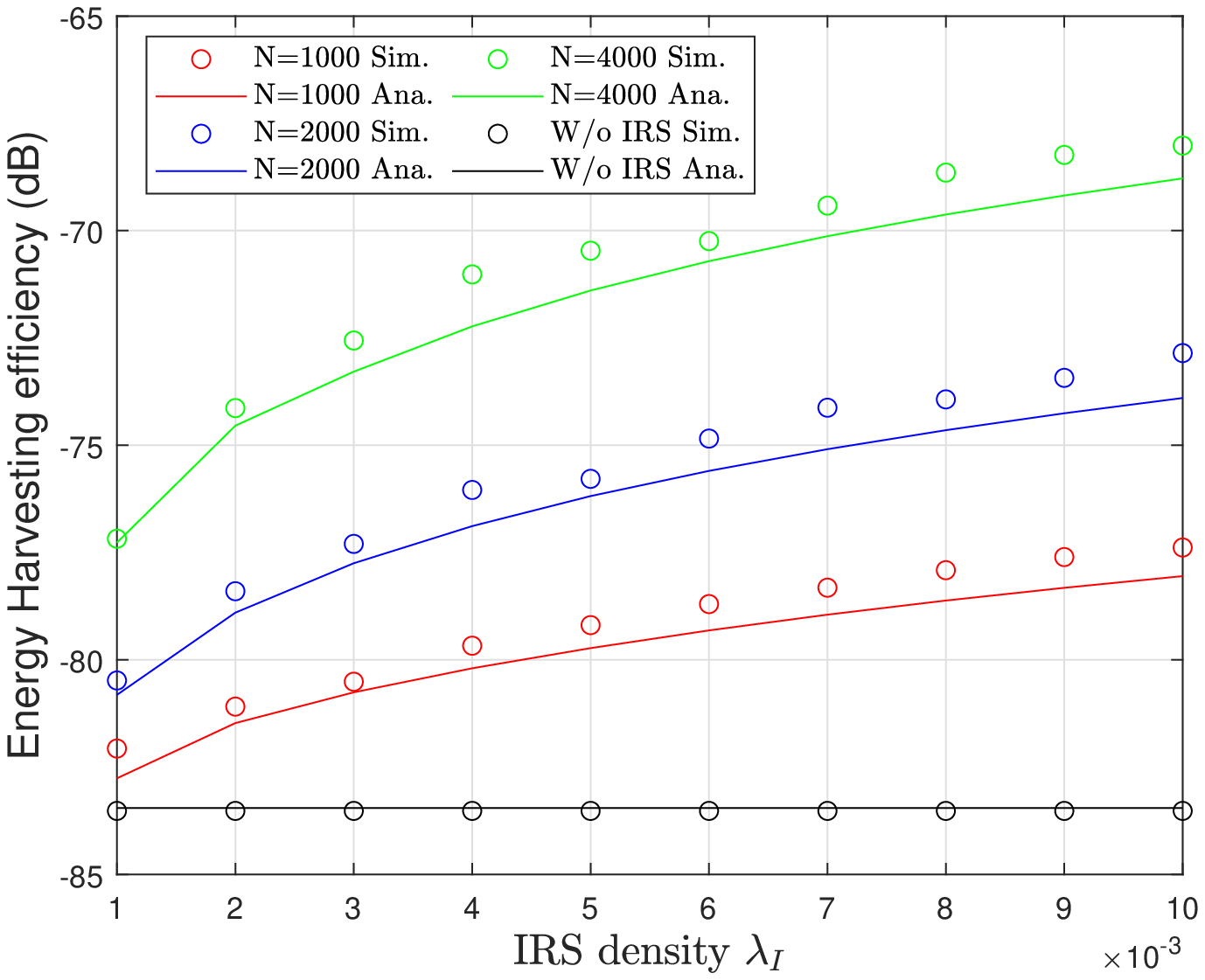}
		\caption{EHE as a function of IRS density $\lambda_{\rm{I}}$ for different IRS reflecting element numbers $N$.}
		\label{ehe}
	\end{minipage}
	\begin{minipage}[t]{0.5\textwidth}
		\includegraphics[scale=0.52]{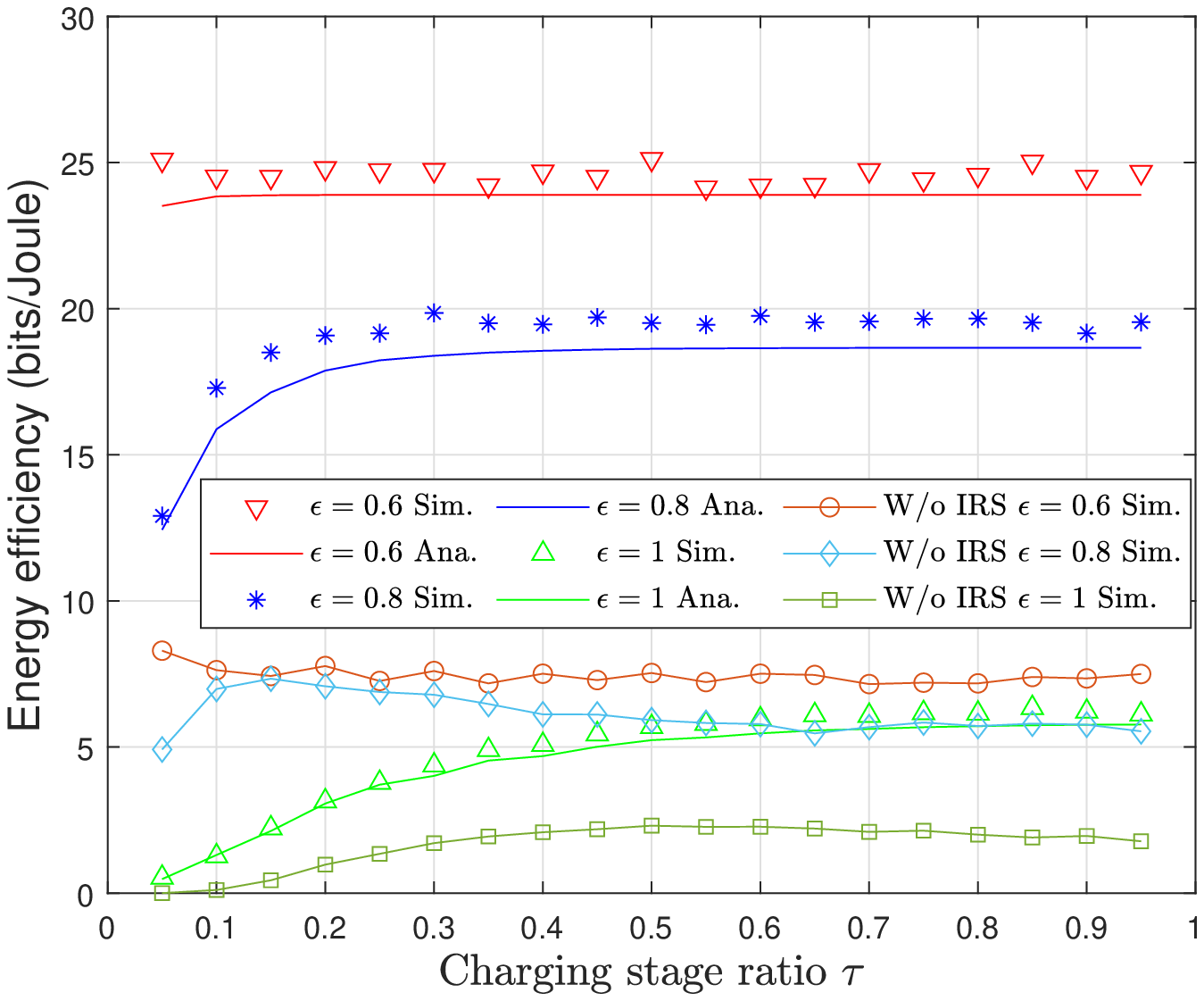}
		\caption{Energy efficiency as a function of charging stage ratio $\tau$ for different power control factors $\epsilon$.}
		\label{ee}
	\end{minipage}
\end{figure}

\textcolor{blue}{In Fig. \ref{ehe}, we evaluate EHE as a function of IRS density $\lambda_{\rm{I}}$ for different IRS reflecting element numbers $N$. We observe that the EHE increases with both $\lambda_{\rm{I}}$ and $N$. However, since the average distance beween an IoT device and its associated BS is as large as 15.8 m, the EHE is small due to the large path loss. Compared to the traditional RF-powered network without IRSs, the higher EHE achieved by the IRS-assisted RF-powered network is credited to the passive beamforming gain. \\ \indent
In Fig. \ref{ee}, we evaluate EE as a function of charging stage ratio $\tau$ for different power control factors $\epsilon$.  We observe that as $\tau$ increases, the EE first increases and then converges to a constant value. A smaller $\epsilon$ results in a higher EE. Due to the power control strategy, the power consumption of an IoT device is mainly dependent on the static power $P_{\text{cu}}$, and thus, the tendency of EE is dominated by the overall coverage probability $\bold{P}_{\text{cov}}$. The variation of EE with $\tau$ can be explained by the tendency of $\bold{P}_{\text{cov}}$ with $\tau$, as discussed in Fig. \ref{wholeprocess}. What's more, referring to Fig. \ref{totalepsilon}, a smaller $\epsilon$ leads to a higher $\bold{P}_{\text{cov}}$, and thus a larger EE. The comparison with the traditional RF-powered network without IRSs shows the effectiveness of IRS beamforming in information transfer. }
\vspace{-1em}
\section{CONCLUSION}
In this work, we evaluated the effect of IRS passive beamforming gain on an IRS-assisted RF-powered IoT network. We adopted the time-switch architecture in which each time slot is partitioned into the charging stage and UL stage, where IRS beamforming was adopted in both EH and UL packet transmission. With the Gamma approximation method, we first characterized the signal power distribution in both charging stage and UL stage, and the interference power distribution in the UL stage. Then, we derived the analytical expressions of energy coverage probability, UL coverage probability, overall coverage probability network spatial throughput \textcolor{blue}{and power efficiency}. By comparing with the conventional RF-powered IoT network without IRSs, we quantified the gain achieved by IRSs in both EH and information transfer. We also optimized the time slot ratio to maximize the network spatial throughput. \textcolor{blue}{There are several interesting concrete future directions to extend this work. One possible direction is to consider the multiple-antenna BS network scenario with the joint optimization of active/passive beamformer design being the focus. Another possible direction is to consider the application of IRSs-assisted RF-powered IoT networks to the mmWave or sub-THz frequency band.}
% or

\appendices
\vspace{-1em}
\section{}
\vspace{-0.5em}
\label{A}
Conditioned on $d_0$ and $x_0$, the Laplace transform of the interference power from other BSs as
\vspace{-0.5em}
\begin{equation}\label{equ63}
	\small
	\begin{aligned}
		&\mathcal{L}_{I|_{d_0,x_0}}\left(s\right)  \triangleq \mathbb{E}\left\{e^{-sI}\right\}\lvert_{d_0,x_0} \approx \mathbb{E}\left\{ e^{-s\left( \sum_{k\in\Lambda_{\rm{u}}'\backslash\left\{0\right\}}{\upsilon_{\rm{I}} P_k f_{\text{d},k}\xi_k}\right)}\right\}\Big|_{d_0,x_0} \\& \overset{\text{(a)}}{=} \mathbb{E}_{\lambda_{\rm{u}}'}\left\{ \prod_{k\in\Lambda_{\rm{u}}'\backslash\left\{0\right\}}{\mathbb{E}_{\xi} \left\{\exp\left(-s\upsilon_{\rm{I}}P_k f_{\text{d},k}\xi_k\right)\right\}}\right\}\Bigg|_{d_0,x_0} \\&\overset{\text{(b)}}{=} \exp\left( -2\pi\lambda_{\rm{u}}'\int_{0}^{\infty}\left(1-e^{-\pi\lambda_{\rm{u}} y^2}\right)\left[1-\mathbb{E}_{\xi,D_k} \left\{ -s\upsilon_{\rm{I}}\rho\left(y^2+H_{\rm{B}}^2\right)^{-\alpha/2} \left(D_k^2+H_{\rm{B}}^2\right)^{\epsilon\alpha/2}\xi_k\right\}\right]y{\rm d}y\right) 
		\\&\overset{\text{(c)}}{=} \exp\left( -2\pi\lambda_{\rm{u}}'\int_{0}^{\infty}\int_0^y{1- \frac{2\pi\lambda_{\rm{u}} D_ke^{-\lambda_{\rm{u}} \pi D_k^2}}{1+s\upsilon_{\rm{I}} \rho\left(y^2+H_{\rm{B}}^2\right)^{-\alpha/2}\left(D_k^2+H_{\rm{B}}^2\right)^{\epsilon\alpha/2}}}{\rm d}D_k y{\rm d}y\right) \\&= \exp\left( -2\pi\lambda_{\rm{u}}'U_I\left(s\upsilon_{\rm{I}}\rho\right)\right),
	\end{aligned}
\end{equation}
where (a) is due to the i.i.d. channel power gain $\xi_k \overset{\text{dist.}}{=}\xi \sim \exp(1),\forall k$ and independent $\Lambda_{\rm{u}}'$, (b) follows from the PGFL of PPP in which the positions of interferers are modeled by a non-homogeneous PPP $\Lambda_{\rm{u}}'$ of density $\lambda_{I_u}(y)=\lambda_{\rm{u}}(1-\exp(-\pi\lambda_{\rm{u}}y^2))$ which is a function of distance relative to the tagged BS, (c) follows from $\mathbb{E}_\xi\left\{e^{-s\xi}\right\} \triangleq \frac{1}{s+1}$ for $\xi \sim \exp(1)$, and $U_I(\cdot)$ is given by
\begin{equation}\label{equ64}
	\small
	\begin{aligned}
		&U_I(x)\triangleq \int_{0}^{\infty}\int_0^y {1- \frac{2\pi\lambda_{\rm{u}} D_ke^{-\lambda_{\rm{u}} \pi D_k^2}}{1+x\left(y^2+H_{\rm{B}}^2\right)^{-\alpha/2}\left(D_k^2+H_{\rm{B}}^2\right)^{\epsilon\alpha/2}}}{\rm d}D_k y{\rm d}y.
	\end{aligned}
\end{equation} 
\vspace{-2.5em}
\section{}
\label{B}
Conditioned on $d_0$ and $x_0$, according to the definition of $\bold{P}_{\rm{en}}$ in (\ref{equ28}), we have
\begin{equation}\label{equ67}
	\small
	\begin{aligned}
		\bold{P}_{\rm{en}}\vert_{d_0,x_0} &\triangleq \mathbb{P}\left\{ \tau T\eta \left(S_{\rm{dr}}+S_{\rm{id}}\right)>E_{\rm{min}}\right\}\vert_{d_0,x_0}=\mathbb{P}\left\{S_{\rm{dr}}+S_{\rm{id}}>C\left( \tau\right)\right\}\vert_{d_0,x_0},
	\end{aligned}
\end{equation}
where $C\left( \tau\right)\triangleq \frac{(1-\tau) \rho x_0^{\epsilon\alpha}}{\tau \eta} $ is a function of $\tau$. \\ \indent
Since either $S_{\rm{dr}}$ or $S_{\rm{id}}$ could be greater than $C(\tau)$, we need to consider the following three cases: $\{S_{\rm{dr}}>C(\tau)\}$, $\{S_{\rm{id}}>C(\tau)\}$ and $\{S_{\rm{dr}}+S_{\rm{id}}>C(\tau)\vert S_{\rm{dr}},S_{\rm{id}}<C(\tau)\}$. It is worth noting that there exists correlation between the events $\left\{S_{\rm{dr}}>C(\tau)\right\}$ and $\left\{S_{\rm{id}}>C(\tau)\right\}$. To simplify the analysis and maintain the tractability, we neglect the correlation and assume that the two events are independent. According to the total probability formula, the conditional energy coverage can be converted to the following form
\vspace{-0.5em}
\begin{equation}\label{equ68}
	\small
	\begin{aligned}
		&\mathbb{P}\left\{S_{\rm{dr}}+S_{\rm{id}}>C\left( \tau\right)\right\}\vert_{d_0,x_0}\approx\mathbb{P}\left\{S_{\rm{dr}}>C(\tau)\right\}\vert_{d_0,x_0}+\mathbb{P}\left\{S_{\rm{id}}>C(\tau)\right\}\vert_{d_0,x_0}\\&+\mathbb{P}\left\{S_{\rm{dr}}+S_{\rm{id}}>C(\tau)\vert S_{\rm{dr}},S_{\rm{id}}<C(\tau)\right\}\vert_{d_0,x_0}\mathbb{P}\left\{S_{\rm{dr}}<C(\tau)\right\}\vert_{d_0,x_0}\cdot\mathbb{P}\left\{S_{\rm{id}}<C(\tau)\right\}\vert_{d_0,x_0}\\&-\mathbb{P}\left\{S_{\rm{dr}}>C(\tau)\right\}\vert_{d_0,x_0}\cdot\mathbb{P}\left\{S_{\rm{id}}>C(\tau)\right\}\vert_{d_0,x_0}.
	\end{aligned}
\end{equation}\indent
We then need to derive the probabilities of the aforementioned three events. Specifically, we have approximated the distribution of $S_{\rm{dr}}\vert_{d_0,x_0}$ by the Gamma distribution $\Gamma\left[k_{\rm{dr}},\theta_{\rm{dr}}\right]$ in (\ref{equ34}). According to Appendix D of \cite{ref21}, for integer $k_{\rm{dr}}$, the probability of $S_{\rm{dr}}>C(\tau)$ conditioned on $x_0$ and $d_0$ is given by
\vspace{-1em}
\begin{equation}\label{equ69}
	\small
	\begin{aligned}
		&\mathbb{P}\left\{S_{\rm{dr}}>C(\tau)\right\}\vert_{d_0,x_0}\approx\frac{\Gamma\left(k_{\rm{dr}},\frac{C(\tau)}{\theta_{\rm{dr}}}\right)}{\Gamma\left(k_{\rm{dr}}\right)}\Bigg|_{d_0,x_0}= \sum_{i=0}^{k_{dr}-1}\frac{\left(-1\right)^i}{i!}\frac{\partial^i}{\partial s^i}\left[\mathcal{L}_{Y_{S\vert_{d_0,x_0}}}\left(s\right)\right]_{s=1},
	\end{aligned}
\end{equation}
where $Y_S\triangleq \frac{C(\tau)}{\theta_{\rm{dr}}}$. Referring to (\ref{equ43}), the Laplace transform of $Y_S$ conditioned on $d_0$ and $x_0$ is derived by
\vspace{-1em}
\begin{equation}\label{equ70}
	\small
	\begin{aligned}
		\mathcal{L}_{Y_S\vert_{d_0,x_0}}\left(s\right) \triangleq \mathbb{E}\left\{e^{-sY_S}\right\}\vert_{d_0,x_0}=\mathcal{L}_{S_{\rm{id}}\vert_{d_0,x_0}}\left(\frac{s}{\theta_{\rm{dr}}}\right),
	\end{aligned}
\end{equation}
where the Laplace transform of $S_{\rm{id}}\vert_{d_0,x_0}$ is given by (\ref{equ63}) in Appendix \ref{A}. \\ \indent
According to the CDF of $S_{\rm{id}}\vert_{d_0,x_0}$ in (\ref{equ45}), the conditional probability of event $\{S_{\rm{id}}>C(\tau)\}$ is given by
\vspace{-1em}
\begin{equation}\label{equ71}
	\small
	\begin{aligned}
		\mathbb{P}\left\{S_{\rm{id}}>C(\tau)\right\}\vert_{d_0,x_0}&=1-\mathbb{P}\left\{S_{\rm{id}}<C(\tau)\right\}\vert_{d_0,x_0}=1-\mathcal{L}^{-1}\left[\frac{1}{s}\mathcal{L}_{S_{\rm{id}}\vert_{d_0,x_0}}\left(s\right)\right]\left(z\right),
	\end{aligned}
\end{equation}
where $z\triangleq C(\tau)$ and $F_{S_{\rm{id}}\vert_{d_0,x_0}}\left(\cdot\right)$ is the conditional CDF of $S_{\rm{id}}$ given by (\ref{equ45}). \\ \indent
We then derive the probability of $\left\{S_{\rm{dr}}+S_{\rm{id}}>C(\tau)\vert S_{\rm{dr}}\right.$ $\left.,S_{\rm{id}}<C(\tau)\right\}$ conditioned on $x_0$ and $d_0$. Because of $S_{\rm{dr}},S_{\rm{id}}<C(\tau)$, we have $C(\tau)-S_{\rm{dr}}>0$ and $C(\tau)-S_{\rm{id}}>0$. According to the approximation $S_{\rm{dr}}\vert_{d_0,x_0}\sim\Gamma\left[k_{\rm{dr}},\theta_{\rm{dr}}\right]$, we have
\vspace{-0.5em}
\begin{equation}\label{equ72}
	\small
	\begin{aligned}
		\mathbb{P}\left\{S_{\rm{dr}}+S_{\rm{id}}>C(\tau)\vert S_{\rm{dr}},S_{\rm{id}}<C(\tau)\right\}\vert_{d_0,x_0} &\approx\mathbb{E}_{S_{\rm{id}}}\left\{\frac{\Gamma\left(k_{\rm{dr}},\frac{C(\tau)-S_{\rm{id}}}{\theta_{\rm{dr}}}\right)}{\Gamma\left(k_{\rm{dr}}\right)}\Bigg| S_{\rm{id}},S_{\rm{dr}}<C(\tau)\right\}\Bigg|_{d_0,x_0} \\& =\sum_{i=0}^{k_{\rm{dr}}-1}{\frac{\left( -1 \right) ^i}{i!}\frac{\partial i}{\partial s^i}\left[ L_{Y_{C\lvert{S_{\rm{dr}},S_{\rm{id}}<C(\tau),d_0,x_0}}}\left( s \right) \right] _{s=1}},
	\end{aligned}
\end{equation}
where $Y_C\triangleq \frac{C(\tau)-S_{\rm{id}}}{\theta_{\rm{dr}}}$ and the Laplace transform of $Y_C$ conditioned on $S_{\rm{dr}},S_{\rm{id}}<C(\tau)$, $d_0$ and $x_0$ is given by
\vspace{-0.5em}
\begin{equation}\label{equ73}
	\small
	\begin{aligned}
		&L_{Y_{C \vert{S_{\rm{dr}},S_{\rm{id}}<C(\tau),d_0,x_0}}}\left(s\right) \triangleq \mathbb{E}\left\{e^{-sY_C}\right\}\vert_{S_{\rm{dr}},S_{\rm{id}}<C(\tau),d_0,x_0} =  \frac{\exp\left(\frac{-sC(\tau)}{\theta_{\rm{dr}}}\right)}{L_{S_{\rm{id}}\vert_{d_0,x_0}}\left(\frac{s}{\theta_{\rm{dr}}}\right)} = \exp\left(V_C\left(s\right)\right),
	\end{aligned}
\end{equation}
where $V_C\left(s\right) \triangleq -\frac{sC(\tau)}{\theta_{\rm{dr}}}+2\pi \lambda _BU\left( \frac{s\upsilon_S}{\theta _{\rm{dr}}} \right)$.  \\ \indent
\vspace{-0.5em}
\section{}
\label{C}
The UL coverage probability conditioned on $d_0$ and $x_0$ is 
\begin{equation}\label{equ77}
	\small
	\begin{aligned}
		\boldsymbol{P}_{\rm{UL}}\vert_{d_0,x_0}&\triangleq\mathbb{P}\left\{\gamma>\overline{\gamma}\right\}\vert_{d_0,x_0}=\mathbb{P}\left\{ S_{\rm{UL}}>{\overline{\gamma}}\left(I + W\right)\right\}\vert_{d_0,x_0} .
	\end{aligned}
\end{equation}
Note that (\ref{equ77}) relates to the conditional SINR distribution through $\bold{P}_{\rm{UL}}\vert_{d_0,x_0} = 1-F_\gamma\left(\overline{\gamma}\right)\vert_{d_0,x_0}$, with $F_\gamma\left(\centerdot\right)\vert_{d_0,x_0}$ being the conditional SINR CDF. \\ \indent
Referring to (\ref{equ46}) in section ${\rm \uppercase\expandafter{\romannumeral3}. A}$, the signal power distribution conditioned on $d_0$ and $x_0$ is approximated by the Gamma distribution $\Gamma\left[ k_{\rm{UL}},\theta_{\rm{UL}}\right]$. Therefore, for integer $k_{\rm{UL}}$, we derive the following conditional UL coverage probability, which is given by
\begin{equation}\label{equ78}
	\small
	\begin{aligned}
		\bold{P}_{\rm{UL}}\vert_{d_0,x_0}&\approx\mathbb{E}_I\left\{\frac{\Gamma\left(k_{\rm{UL}},\frac{\overline{\gamma}\left(I+W\right)}{\theta_{\rm{UL}}}\right)}{\Gamma\left(k_{\rm{UL}}\right)}\right\}\Bigg|_{d_0,x_0} = \sum_{i=0}^{k_{\rm{UL}}-1}{\frac{\left(-1\right)^i}{i!}\frac{\partial^i}{\partial s^i}\left[\mathcal{L}_{Y_I\vert_{d_0,x_0}}\left(s\right)\right]_{s=1}},
	\end{aligned}
\end{equation}
where $\Gamma(\cdot,\cdot)$ denotes the upper incomplete Gamma function, and $Y_I\triangleq \frac{\overline{\gamma}\left(I+W\right)}{\theta_{\rm{UL}}}$. Based on (\ref{equ63}) in Appendix \ref{A}, the
Laplace transform of $Y_I$ conditioned on $d_0$ and $x_0$ is given by
\begin{equation}\label{equ79}
	\small
	\begin{aligned}
		\mathcal{L}_{Y_I\vert_{d_0,x_0}}\left(s\right)&\triangleq\mathbb{E}\left\{e^{-sY_I}\right\}\vert_{d_0,x_0} = \exp\left(-\frac{s \overline{\gamma}W}{\theta_{\rm{UL}}}\right)\mathcal{L}_{I\lvert_{d_0,x_0}}\left( \frac{s\overline{\gamma}}{\theta_{\rm{UL}}}\right) = \exp\left(V\left(s\right)\right),
	\end{aligned}
\end{equation}
where \( V\left(s\right)\triangleq-\frac{s\overline{\gamma}W}{\theta_{\rm{UL}}}-2\pi\lambda_{\rm{u}}'U_I\left(\frac{s\overline{\gamma}\upsilon_{\rm{I}}\rho}{\theta_{\rm{UL}}}\right)\). 
% if you want by leaving the argument blan



%\ifCLASSOPTIONcaptionsoff
%  \newpage
%\fi



%\bibliographystyle{IEEEtran}
\footnotesize
\bibliographystyle{IEEEtran}
\bibliography{cited}

%\begin{thebibliography}{1}
%\end{thebibliography}

% \includegraphics command within an optional argument. (You coULd create

%\end{IEEEbiography}


%\newpage



%\vfill

% Can be used to pULl up biographies so that the bottom of the last one

\end{document}